\documentclass[a4paper,fleqn]{cas-dc}

\usepackage[authoryear]{natbib}
\usepackage{amsmath,amssymb,amsfonts}
\usepackage{algorithmic}
\usepackage{graphicx}
\usepackage{textcomp}
\usepackage{xcolor}
\usepackage{ragged2e}
\usepackage{colortbl}
\usepackage{multirow}
\usepackage{balance}
\usepackage[figuresright]{rotating}
\usepackage{rotfloat}
\usepackage{color}
\usepackage{hyperref}

\usepackage{url}
\usepackage{caption}
\usepackage{graphicx}
\usepackage{float} 
\usepackage{subcaption}
\usepackage{balance}
\usepackage{hyperref}

\usepackage{utfsym}
\usepackage{listings}
\usepackage{xcolor}
\usepackage{colortbl}
\definecolor{codegreen}{rgb}{0,0.6,0}
\definecolor{codegray}{rgb}{0.5,0.5,0.5}
\definecolor{codeorange}{rgb}{1,0.49,0}
\definecolor{backcolor}{rgb}{0.95,0.95,0.96}

\usepackage{tcolorbox}
\usepackage{colortbl}
\usepackage{framed} 
\lstdefinestyle{mystyle}{
 backgroundcolor=\color{backcolor},
 commentstyle=\color{codegray},
 keywordstyle=\color{codeorange},
 numberstyle=\tiny\color{codegray},
 stringstyle=\color{codegreen},
 basicstyle=\ttfamily\footnotesize,
 breakatwhitespace=false,
 breaklines=true,
 captionpos=b,
 keepspaces=true,
 numbers=left,
 numbersep=5pt,
 showspaces=false,
 showstringspaces=false,
 showtabs=false,
 tabsize=2,
 xleftmargin=10pt,
}
\begin{document}
\begin{sloppypar}

\let\WriteBookmarks\relax
\def\floatpagepagefraction{1}
\def\textpagefraction{.001}

\shorttitle{Automated Detection of Inter-Language Design Smells in Multi-Language Deep Learning Frameworks}
\shortauthors{Z Li et al.}
\title [mode = title]{Automated Detection of Inter-Language Design Smells in Multi-Language Deep Learning Frameworks}
    
\author[1]{Zengyang Li}
\ead{zengyangli@ccnu.edu.cn}

\credit{Conceptualization of this study, Methodology, Investigation, Data curation, Writing - Original draft preparation}
\address[1]{School of Computer Science \& Hubei Provincial Key Laboratory of Artificial Intelligence and Smart Learning, \\Central China Normal University, Wuhan, China \\}

\author[1]{Xiaoyong Zhang}
\ead{charles@mails.ccnu.edu.cn}
\credit{Conceptualization of this study, Methodology, Investigation, Data curation, Software, Writing - Original draft preparation}

\author[1]{Wenshuo Wang}
\ead{wenshuowang@mails.ccnu.edu.cn}
\credit{Investigation, Data curation, Software, Writing - Original draft preparation}

\author[2]{Peng Liang}
\cormark[1]
\ead{liangp@whu.edu.cn}
\credit{Conceptualization of this study, Methodology, Writing - Original draft preparation}
\address[2]{School of Computer Science, Wuhan University, Wuhan, China}

\author[1]{Ran Mo}
\ead{moran@ccnu.edu.cn}
\credit{Methodology, Writing - Original draft preparation}

\author[3]{Jie Tan}
\ead{j.tanjie@outlook.com}
\credit{Methodology, Writing - Original draft preparation}
\address[3]{Intelligent Game and Decision Lab, Beijing, China}

\author[4]{Hui Liu}
\ead{hliu@hust.edu.cn}
\credit{Methodology, Writing - Original draft preparation}
\address[4]{School of Artificial Intelligence and Automation, Huazhong University of Science and Technology, Wuhan, China}

\cortext[cor1]{Corresponding author.}

\begin{abstract}
\noindent \textbf{Context}: 
Nowadays, most deep learning frameworks (DLFs) use multilingual programming of Python and C/C++, facilitating the flexibility and performance of the DLF. However, inappropriate inter-language interaction may introduce design smells involving multiple programming languages (PLs), i.e., Inter-Language Design Smells (ILDS). Despite the negative impact of ILDS on multi-language DLFs, there is a lack of an automated approach for detecting ILDS in multi-language DLFs and a comprehensive understanding on ILDS in such DLFs.\\ 
\noindent\textbf{Objective}: This work aims to automatically detect ILDS in multi-language DLFs written in the combination of Python and C/C++, and to obtain a comprehensive understanding on such ILDS in DLFs. \\
\noindent\textbf{Methods}: We first developed an approach to automatically detecting ILDS in the multi-language DLFs written in the combination of Python and C/C++, including a number of ILDS and their detection rules defined based on inter-language communication mechanisms and code analysis. Then, we developed the \textsc{CPsmell} tool that implements detection rules for automatically detecting such ILDS, and manually validated the accuracy of the tool. Finally, we performed an empirical study to evaluate the ILDS in multi-language DLFs.      \\
\noindent\textbf{Results}: We proposed seven ILDS and achieved an accuracy of 98.17\% in the manual validation of \textsc{CPsmell} in 5 popular multi-language DLFs. The study results revealed that among the 5 DLFs, TensorFlow, PyTorch, and PaddlePaddle exhibit relatively high prevalence of ILDS; each smelly file contains around 5 ILDS instances on average, with ILDS \textit{Long Lambda Function For Inter-language Binding} and \textit{Unused Native Entity} being relatively prominent; throughout the evolution process of the 5 DLFs, some ILDS were resolved to a certain extent, but the overall count of ILDS instances shows an upward trend. \\
\noindent\textbf{Conclusions}: The automated detection of the proposed ILDS achieved a high accuracy, and the empirical study provides a comprehensive understanding on ILDS in the multi-language DLFs. 
\end{abstract}


\maketitle

\section{Introduction}
Deep learning (DL) technology, as a branch of machine learning (ML), stands out as a prominent manifestation of artificial intelligence (AI) development, finding extensive applications in fields such as commerce, healthcare, and industry. One of the core elements behind this technology is the development of its algorithms and models. Traditionally, developing and fine-tuning these algorithms and models was a complex and time-consuming task. However, the advent of deep learning frameworks (DLFs) has substantially simplified this process, providing researchers and engineers with more efficient and intuitive tools. Nevertheless, as DL technology evolves, ensuring the maintainability of DLFs remains a challenge. High maintainability not only prolongs the lifespan of DLFs but also ensures their consistent and reliable performance over time~\citep{yang2022comprehensive}.

Software design smells refer to poor programming practices that can negatively impact the software quality despite meeting functional requirements in the software development process~\citep{moha2009decor, abidi2021multi}. These usually include code smells~\citep{fowler2018refactoring}, which manifest as potential issues at the code level, and anti-patterns, which represent higher-level design flaws that may span architecture, design, or project management. These design smells have a direct negative impact on software maintainability. They can make the code more difficult to understand, modify, and extend, increasing maintenance costs and introducing potential errors. Therefore, avoiding such design smells and adopting best practices are of paramount importance.

Several studies have been conducted on design smells in ML software, and they focus on only one PL, i.e., Python~\citep{gesi2022code,  zhang2022code, van2021prevalence, jebnoun2020scent}. However, many DLFs are implemented by two or more PLs. Each PL possesses unique features and functionalities, and certain PLs may excel in particular tasks. Focusing on just one PL may disregard the complexities of interactions between multiple PLs, encompassing challenges related to multi-language linking and dependencies. Additionally, diverse PLs use varied approaches to address security concerns. Overlooking these distinctions can potentially expose systems to increased security vulnerabilities~\citep{li2022polycruise}. Moreover, codes in various PLs are intertwined in the system. If these codes cannot be effectively divided into modules to ensure safe and efficient operations, it will make the code difficult to understand and maintain
~\citep{grichi2020towards, kullbach1998program, mayer2017multi}.

Python and C/C++ are often used for multi-language programming in DLFs~\citep{grichi2020MLF}, such as PyTorch. Due to the simplicity and usability of Python, their upper-layer interfaces are often written in Python, allowing users to create DL models more easily, while the underlying core computing part is mainly written in C++ because C++ can provide efficient memory management and computing performance. Although there are many benefits of using multi-language programming, the increase of PLs will also lead to more design smells~\citep{grichi2021impact} and bugs in the software~\citep{kochhar2016large, yang2023demystifying, yang2024multi}. 
These design smells occur only in multi-language programming and  involve interactions between multiple PLs.
They cannot be identified by single-language smell detection tools. We refer to these design smells involving interactions between PLs as inter-language design smells (ILDS\footnote{ILDS can be singular or plural depending on the context.}).

In this work, we conduct an in-depth analysis of ILDS in popular multi-language DLFs. The main contributions of this work are summarized as follows:
\begin{itemize}
\item[$\bullet$]We proposed 7 ILDS with detection rules in multi-language DLFs written in the combination of Python and C/C++.
\item[$\bullet$] We developed the \textsc{CPsmell} tool to implement the detection rules for the 7 ILDS, and made a manual validation of the tool in 5 popular multi-language DLFs, achieving an accuracy of 98.17\%. 
\item[$\bullet$] We conducted an empirical study in the 5 DLFs to evaluate their maintainability in terms of the proposed ILDS . 
\end{itemize}

The rest of this paper is organized as follows. Section \ref{sec_relatedwork} discusses the work related to our research, Section \ref{sec_ILDS} presents 7 ILDS that we proposed, Section \ref{sec_tool} implements and validates the ILDS detection, Section \ref{sec_empiricalstudy} reports the empirical study on ILDS in DLFs, Section \ref{sec_limitation} presents the limitations of this work, Section \ref{sec_implication} provides implications for practitioners and researchers, and Section \ref{sec_conclusion} concludes this work with future research directions.

\section{RELATED WORK}\label{sec_relatedwork}

\textbf{Multi-language software quality.} 
Existing studies on the quality of multi-language software systems mainly investigate the bugs and vulnerabilities in such systems. Hu \textit{et al}. studied the evolution of Python/C API and summarized 10 bug patterns~\citep{hu2020python}. Monat and Ouadjaout reported runtime errors that may occur in Python, C, and their language interfaces~\citep{monat2021multilanguage}. Li \textit{et al}. conducted an empirical study on bugs in three multi-language DLFs, i.e., MXNet, PyTorch, and TensorFlow, analyzing the characteristics of the bugs and their impact on the development of DLFs ~\citep{li2023DLF}. Li \textit{et al}. analyzed 66,932 bugs from 54 Apache projects and studied the characteristics of bug resolution in multi-language software systems~\citep{li2023understanding}. Youn \textit{et al}. extended the static code analysis tool CodeQL to track the inter-language boundary data flow of Java-C and Python-C, and identified interoperation bugs in two multi-language programs~\citep{youn2023declarative}. Li \textit{et al}. defined four types of inter-language interfaces and developed the PolyFax tool for the identification of inter-language interfaces and the analysis of vulnerabilities in commit messages~\citep{li2022polyfax, li2022vulnerability, li2024multilingual}. The aforementioned studies address two external quality attributes, i.e., reliability and security, while \textbf{our study} focuses on the maintainability, a key internal quality attribute of multi-language software.

\textbf{Design smells in ML software.} 
There have been several studies on design smells in ML applications, and most of the studies focus only on Python.
Gesi \textit{et al}. identified five code smells (e.g., scattered use of ML Library) unique to DL-specific code~\citep{gesi2022code}. Jebnoun \textit{et al}. performed a comparative study on ten code smells (e.g., long method) in the Python code of DL and traditional non-DL applications, and found that code smell instances in DL code are more complex ~\citep{jebnoun2020scent}. 
Van Oort \textit{et al}. investigated the prevalence of code smells in 74 open-source ML projects in Python, and found that code duplication is widespread~\citep{van2021prevalence}. Zhang \textit{et al}. collected 22 ML-specific code smells when using ML-related Python libraries~\citep{zhang2022code}. These studies explored smells of ML software in Python. In contrast, \textbf{our work} investigates design smells in DLFs in the context of multi-language programming. 



\textbf{Inter-language design smells (ILDS).} There are a few studies on the design smells rooted in the interaction between multiple PLs. Abidi \textit{et al}. conducted research on document and multi-language system retrieval using keywords, where they identified 12 design smells~\citep{abidi2019code} and 6 anti-patterns~\citep{abidi2019anti}. They subsequently proposed an ILDS detection tool for multi-language systems written in the combination of Java and C/C++, and detected 15 ILDS in 98 versions of 9 open-source JNI projects~\citep{abidi2021multi}. Groot \textit{et al}. elicited eight general unexpected dependencies within ASML's multi-language software systems by interviewing 17 practitioners from six different software development teams in ASML~\citep{groot2024catalog}. In contrast, \textbf{our study} proposed ILDS in multi-language DLFs written in Python and C/C++, implemented a dedicated detection tool for such ILDS, and validated the ILDS with detection rules in popular DLFs.

\textbf{Inter-language communication mechanisms (ILCMs).} 
Previous studies on multi-language software systems written in the combination of Python and C/C++ have mostly relied on an ILCM known as the Python/C API~\citep{hu2020python, monat2021multilanguage, youn2023declarative}. However, the ILCMs used by different multi-language software systems vary significantly, and a single software system may employ multiple ILCMs. By reading the source code of multiple popular DLFs, \textbf{our study} identified three ILCMs that are commonly used in DLFs, namely Python/C API~\citep{python/c_api}, pybind11~\citep{pybind11} and ctypes~\citep{ctypes}. Moreover, we developed the \textsc{CPsmell} tool for ILDS detection, which is capable of identifying and parsing the three ILCMs.

\section{INTER-LANGUAGE DESIGN SMELLS}\label{sec_ILDS}
To identify ILDS in multi-language DLFs written in Python and C/C++, first, we examined diverse sources, including Stack Overflow, technical documentation, literature, and GitHub issues. Second, using keywords like ``deep learning'', ``Python/C API'', ``pybind'', ``ctypes'', ``SWIG'', ``CFFI'', ``Cython'', ``multi-language'', ``cpp-extensions'', ``language interface'', ``polyglot programming'' and ``language interaction'', we performed relevant searches and mapped the identified issues to the source code of DLFs to extract ILDS. Specifically, we focused on popular DLFs, such as TensorFlow and PyTorch.
For an identified ILDS, we followed two screening criteria: First, the ILDS must be identified based on at least two information sources, with at least one being authoritative (e.g., official documentation);
Second, we ensured that each ILDS could be found in the source code of at least two DLFs.

\begin{figure*}[h]
  \centering
  \includegraphics[width=0.7\textwidth]{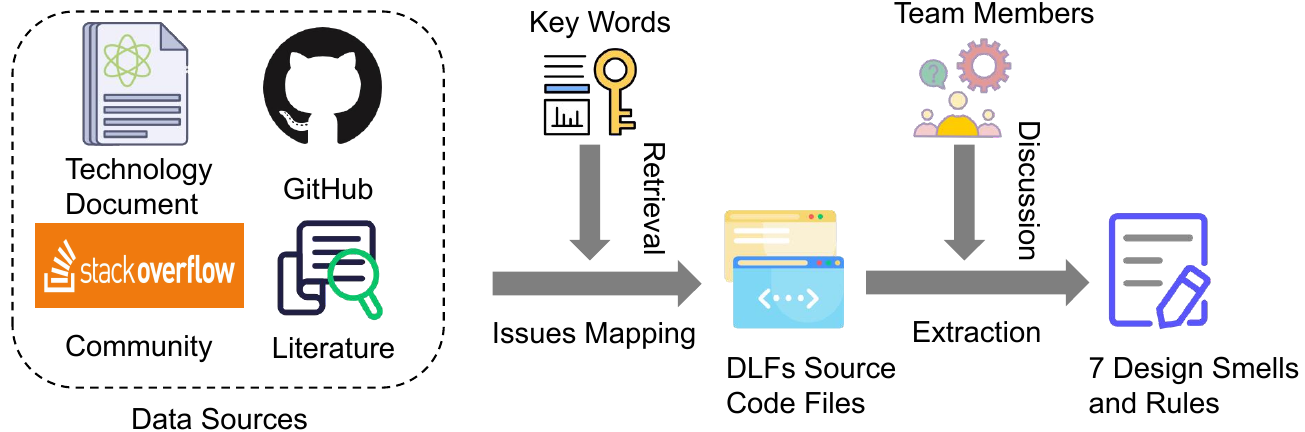}
  \caption{Inter-language design smell extraction process}\label{fig:f1}
\end{figure*}

Finally, through team discussions, we identified 7 ILDS in DLFs and summarized an identification rule for each ILDS. 
The 7 ILDS are \emph{Unused Native Entity (UNE)}, \emph{Long Lambda Function For Inter-language Binding (LLF)},  \emph{Lack of Rigorous Error Check (LREC)}, \emph{Lack of Static Declaration (LSD)}, \emph{Not Using Relative Path (NURP)}, \emph{Large Inter-language Binding Class (LILBC)}, \emph{Excessive Inter-Language Communication (EILC)}. Among these, UNE, LLF, LREC, LSD, and NURP are at the code level and considered as code smells, while LILBC and EILC are at the architecture level and regarded as anti-patterns.
Figure~\ref{fig:f1} summarizes the entire extraction process for ILDS. In the following, we will provide specific descriptions of the 7 ILDS.

\subsection{Unused Native Entity (UNE)} 

\textbf{Description: }When using Python/C API or pybind11 to extend Python with C/C++, it is necessary to create an extension module written in C/C++, which is similar to a Python module. The extension module is bound with classes, functions, and variables written in C/C++. After compiling the extension module, it can be imported into Python, enabling interaction between the two languages. We summarize these C/C++ codes for interaction as native entities. Depending on the granularity of the code, we primarily consider three types of entities: modules, classes, and functions. Furthermore, since functions in Python can exist independently of classes, we consider the functions and classes at the same level of granularity only but neglect the functions within unused classes.

An extension module is written in C/C++ and can bind various entities for interaction, including classes, functions, and variables. When an extension module is used, it does not mean that all entities in this module are used. Only when every entity in an extension module is not used, this module is considered as an unused module. To ensure the independence of data of different entities, when we deem a module unused, we do not consider which specific entities within it are unused.

\textbf{Impact: } If developers do not have a comprehensive understanding of the entire system, to avoid destroying its relevance, such \textit{UNE} will remain in the software system for a long time. Maintainers may be confused by these \textit{UNE} because they do not realize whether these native entities are used. Consequently, when updating or upgrading the system, a review of these entities and their dependencies becomes necessary, which increases the complexity of development work.

\textbf{Refactoring Recommendation: }Remove the unused native code associated with these entities or add deprecated annotations.

\textbf{Identification Method: }
\begin{equation}\tag{Rule1}\label{eq1}
\footnotesize{
\begin{gathered}
m:=ExtensionModule \\
f:=NativeFunction \\
c:=NativeClass \\
IsBound(f,c)=True\  \textbf{AND}\ IsCalled(f,c)=False\ \textbf{AND}\ \\ IsImported(m,f,c)=False\ \textbf{OR}\\ \   InBUILD(m)=True\ \textbf{AND}\ IsLoaded(m)=False\ 
\end{gathered}
}
\end{equation}
We define an extension module $m$, which represents a C/C++ extension module written using Python/C API or pybind11. A native function $f$ represents a C/C++ function.
A native class $c$ represents a C++ class. 

For the module \textit{m}, this rule detects \textit{m} that is not imported into Python and has not been loaded through \textsc{LoadLibrary}. Additionally, this rule checks for extension modules that are imported into Python via configuration in a BUILD file. For the class \textit{c} and function \textit{f}, this rule identifies situations where they are bound but are neither called nor imported by Python.

\textbf{Example: }As shown in Listing~\ref{lst:code1}, in PyTorch version 1.11, the \textsc{test\_deploy\_lib.cpp} file used pybind11 to define an extension module named \textsc{libtest\_deploy\_lib}~\citep{pytorch_uem}. This module bounds 9 C++ functions for use in Python. However, no Python files use this extension module. In PyTorch versions 2.0.1 and later, this extension module and the 9 C++ functions were removed.
\begin{lstlisting}[language=C++,caption=Design Smell - Unused Native Entity,label={lst:code1},mathescape=true,breaklines=true]
PYBIND11_MODULE(libtest_deploy_lib, m) {
  m.def("raise_and_catch_exception", raise_and_catch_exception);
  m.def("raise_exception", raise_exception);
  m.def("check_initial_state", check_initial_state);
  m.def("get_in_another_module", get_in_another_module);
  m.def("set_in_another_module", set_in_another_module);
  m.def("get_bar", get_bar);
  m.def("set_bar", set_bar);
  m.def("get_bar_destructed", get_bar_destructed);
  m.def("simple_add", simple_add);
}
\end{lstlisting}

\subsection{Long Lambda Function For Inter-language Binding (LLF)}

\textbf{Description: }In pybind11, \textsc{m.def} is used for module-level function bindings. It primarily takes two parameters: one is the name of the function to be made available in Python, and the other is the corresponding C/C++ function. In the development of DLFs, developers often use Lambda expressions to represent the second parameter of \textsc{m.def}. This approach avoids the need to define a separate new function, making the program more flexible and the code more concise. However, some Lambda expressions can become excessively long, leading to the \textit{LLF}. Chen~\citep{chen2016detecting} and Jebnoun~\citep{jebnoun2020scent} have both studied the use of long Lambda functions as a bad smell in the Python code only. In contrast, we focus on the use of Lambda functions when C/C++ is bound to Python. Moreover, the binding code requires the use of special macro definitions, making it challenging for single-language code smell detection tools to analyze.

\textbf{Impact: }An excessively long Lambda expression makes the binding of the function look complicated, which not only violates the original design intention of Lambda expressions, but also affects the readability of the program and the maintainability of the system.

\textbf{Refactoring Recommendation: }In the file of the extension module, we can define a dedicated C/C++ function, and associate it with Python using its function reference during the binding process.

\textbf{Identification Method: }
\begin{equation}\tag{Rule2}\label{eq2}
\footnotesize{
\begin{gathered}
    IsLF(f)=True \ \textbf{AND} \ Length(f)>MaxLengthOfLFThreshold
\end{gathered}
}
\end{equation}
The native function $f$ is defined in~\ref{eq1}. 
This rule identifies the situation where $f$ is a Lambda function and the length of $f$ exceeds the threshold set for the length of the Lambda function. We have adopted the default threshold value of 80 characters as established by Chen~\citep{chen2016detecting}. The basis for this decision is that Google's style guide for Python code~\citep{google_style} suggests that the length of a Lambda function should ideally not exceed 80 characters, which aligns with our usage scenario.

\textbf{Example: }As shown in Listing~\ref{lst:code2}, this segment of code is a binding using pybind11~\citep{tensorflow_llf}, where \textsc{TFE\_GetTaskStates} is the name of the Python function, and the subsequent lambda function serves as the native implementation of \textsc{TFE\_GetTaskStates}. This lambda function has reached 50 lines, making it quite bulky in the process of binding.
\begin{lstlisting}[language=C++,mathescape=true,breaklines=true,firstnumber=1]
m.def("TFE_GetTaskStates", [](py::handle& ctx,
    const std::vector<std::string>& job_names,
    const std::vector<int>& task_nums) {
\end{lstlisting}
{\lstset{aboveskip=0pt,belowskip=0pt} 
\begin{lstlisting}[language=C++,numbers=none]
...
\end{lstlisting}
}
\begin{lstlisting}[language=C++,caption=Design Smell - Long Lambda Function For Inter-language Binding,label={lst:code2},mathescape=true,breaklines=true,firstnumber=45]
        output[i] = py::none(); } }
    tensorflow::MaybeRaiseRegisteredFromTFStatus(status.get());
    return tensorflow::PyoOrThrow(output.release().ptr());  });
\end{lstlisting}


\subsection{Lack of Rigorous Error Check (LREC)} \label{sec_LREC}

\begin{figure*}[h]
  \centering
  \includegraphics[width=0.8\textwidth]{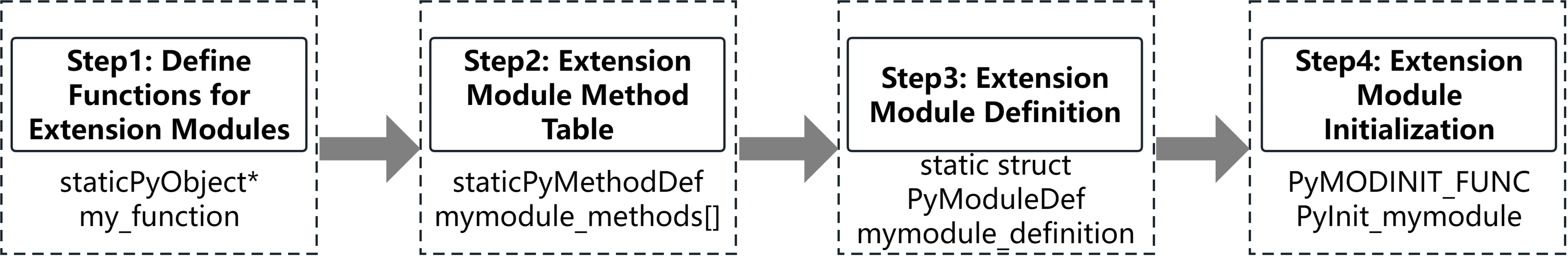}
  \caption{Process of creating an extension module using the Python/C API}\label{fig:f2}
\end{figure*}

\textbf{Description: }The process of creating extension modules using the Python/C API is shown in Figure~\ref{fig:f2}. Throughout this process, developers typically write a substantial amount of code without any error detection. This is especially noticeable during the initialization of a module.

\textbf{Impact: }During the development of DLFs, developers usually use function \textsc{PyModule\_AddObject()} to add an object to the module during module initialization. This function will steal the reference count of an object, but it is different from other functions that steal reference counts; if \textsc{PyModule\_AddObject()} fails to add an object, the reference count of the object will not be decreased. Frequent use of this function can easily cause memory leaks. Moreover, if the return value of this function is not properly handled, developers may be left unaware of the specific source of errors when problems arise in the program. They typically resort to traversing through all the code associated with module creation, which increases the complexity of development work. Therefore, it is necessary to implement rigorous error detection for this function.

\textbf{Refactoring Recommendation: }According to the official Python documentation~\citep{python/c_api_lrec}, it is recommended to use the new \textsc{PyModule\_AddObjectRef()} in Python 3.10 instead of \textsc{PyModule\_AddObject()}. The \textsc{PyModule\_AddObjectRef()} function will not steal the reference count of the object regardless of whether it is successful or not, which facilitates memory management. Subsequently, it is advisable to perform an error check on the return value of this function, manually throw exceptions, or use \textsc{try/except} block to systematically capture and handle exceptions.

\textbf{Identification Method: }
\begin{equation}\tag{Rule3}\label{eq3}
\footnotesize{
\begin{gathered}
x:=f(y)|f\in \{GetPyModule\_AddObjectField\} \\
CheckReturnValue(x)=False\ \textbf{AND}\ \\ InTryExcept(x)=False
\end{gathered}
}
\end{equation}
$x$ represents a series of operations in the extension module creation process, including adding a reference to an object using the \textsc{PyModule\_AddObject} function during module initialization. This rule checks the return value of $x$ and identifies whether $x$ exists in the \textsc{try/except} code block.

\textbf{Example: }\textsc{THPSize\_init} is a module in PyTorch related to initialization of the tensor size. Listing~\ref{lst:code4-1} illustrates its code structure in version 0.3.1~\citep{pytorch_lrec0.3.1}. The \textsc{PyModule\_AddObject()} function adds an object named \textsc{Size} to this module, but this process lacks rigorous error checking. In PyTorch version 1.1.0~\citep{pytorch_lrec1.1.0}, the module underwent refactoring, as shown in Listing~\ref{lst:code4-2}, which includes a check on the return value of the \textsc{PyModule\_AddObject()} function.
\begin{lstlisting}[language=C++,caption=Design Smell - Lack of Rigorous Error Check,label={lst:code4-1},mathescape=true,breaklines=true]
bool THPSize_init(PyObject *module)
{
  THPSizeClass = (PyObject*)&THPSizeType;
  if (PyType_Ready(&THPSizeType) < 0)
    return false;
  Py_INCREF(&THPSizeType);
  PyModule_AddObject(module, "Size", (PyObject *)&THPSizeType);
  return true;
}
\end{lstlisting}
\begin{lstlisting}[language=C++,caption=Refactoring - Lack of Rigorous Error Check,label={lst:code4-2},mathescape=true,breaklines=true]
void THPSize_init(PyObject *module)
{
  if (PyType_Ready(&THPSizeType) < 0) {
    throw python_error();
  }
  Py_INCREF(&THPSizeType);
  if (PyModule_AddObject(module, "Size", (PyObject*)&THPSizeType) < 0) {
    throw python_error();
  }
}
\end{lstlisting}

\subsection{Lack of Static Declaration (LSD)}

\textbf{Description: }When writing extension modules in C/C++ using the Python/C API, it is advisable to declare C/C++ functions as static to prevent naming conflicts with functions in other extension modules, which can lead to namespace pollution. However, in certain scenarios, developers may wish to make these functions available for use by other extension modules, and thus declare them as non-static. In fact, according to the official Python documentation~\citep{python/c_api_lsd} and the instructions of Rossum~\citep{rossum2018extending}, all modules can work normally when they are statically linked to the Python interpreter. For static functions that are required to be called by other extension modules, the Capsule mechanism in C++ can be employed. In practical development projects, faced with a diverse range of C/C++ extension functions, developers may ignore this aspect, leading to this design smell.

\textbf{Impact: }In future system extensions and upgrades, if newly added functions share the same name with existing functions in an extension module and these shared-name functions are not declared as static, conflicts may arise when binding Python functions with C/C++ functions. The interpreter will not distinguish which C/C++ function to use, potentially leading to linkage errors. This situation is not easily detectable, but can result in unpredictable harm to the system, increasing the workload of developers. These nonstatic C/C++ functions can significantly impact the portability and scalability of the entire system.

\textbf{Refactoring Recommendation: }Add the static declaration to native C/C++ functions that are not declared static. We recommend that developers strictly follow the naming conventions. 

\textbf{Identification Method: }
\begin{equation}\tag{Rule4}\label{eq4}
\footnotesize{
\begin{gathered}
IsBound(f)=True \ \textbf{AND}\ IsImplemented(f)=True\ \textbf{AND}\ \\ IsDeclaredStatic(f)=False
\end{gathered}
}
\end{equation}
When a C/C++ extension module is written using the Python/C API, if a native function \textit{f} bound in the module is not declared static during the implementation process, this ILDS will be introduced.

\textbf{Example: }As shown in Listing~\ref{lst:code5}, \textsc{THPModule\_setDefaultDtype} is a native C++ function. Its purpose is to set a default data type for PyTorch tensors~\citep{pytorch_lsd}. If this module is ported, or in the future, if new extension modules have the same requirement, and developers might not be aware of this function, it could lead to naming conflicts.
\begin{lstlisting}[language=C++,caption=Design Smell - Lack of Static Declaration,label={lst:code5},mathescape=true,breaklines=true]
PyObject* THPModule_setDefaultDtype(PyObject* _unused, PyObject* dtype) {
  HANDLE_TH_ERRORS
  torch::tensors::py_set_default_dtype(dtype);
  Py_RETURN_NONE;
  END_HANDLE_TH_ERRORS
}
\end{lstlisting}

\subsection{Not Using Relative Path (NURP)} 

\textbf{Description: }In multi-language systems written in Python and C/C++ , if developers just want to call a common C/C++ function or variable, ctypes is a good choice. ctypes is a Python library that provides a way to call dynamic link libraries (DLLs) and enables Python to interact with C/C++. This ILDS occurs when developers specify the DLL to load simply by a name or by an absolute path when using ctypes.

\textbf{Impact: }A large software system is often completed by multiple teams, and some developers may not fully understand the architecture of the entire software system. When loading DLLs using only their names, the operating system searches for these DLLs in predefined paths. This loading process can be perplexing to other developers or system maintainers who are unaware of the specific locations of these DLLs in the operating system~\citep{stackoverflow}, which affects code portability. We found 4092 questions with the ctypes tag on Stack Overflow, 52\% of which were related to DLLs. If this ILDS exists, the unspecified library path will make debugging the system more difficult when a DLL loading problem happens. 

\textbf{Refactoring Recommendation: }Try to avoid using just one name when loading DLLs. The path to the library should be clearly defined or use a relative path.

\textbf{Identification Method: }
\begin{equation}\tag{Rule5}\label{eq5}
\footnotesize{
\begin{gathered}
x:=f(y)|f\in \{GetcdllField, GetCDLLField, GetLoadlibraryField, \\ GetwindllField, GetWinDLLField\} \\
IsImported(ctypes)=True\ \textbf{AND}\  LoadDLL(
x)=True\ \textbf{AND}\ \\ RelativePath(x)=False
\end{gathered}
}
\end{equation}
The rule identifies the situation when ctypes is imported and the DLL is loaded without using a relative path. $x$ represents several ways to load the DLL.

\textbf{Example: }Listing~\ref{lst:code6} shows that PyTorch needs to load the NVML library when obtaining information from the NVidia GPU. We found that the developer only gave the name of the DLL without indicating its path when implementing this operation~\citep{pytorch_nurp}. Additionally, there was no restriction on the operating system used for this operation.
\begin{lstlisting}[language=Python,caption=Design Smell - Not Using Relative Path,label={lst:code6},mathescape=true,breaklines=true]
from ctypes import CDLL, c_int, byref
    nvml_h = CDLL("libnvidia-ml.so.1")
\end{lstlisting}

\subsection{Large Inter-language Binding Class (LILBC)}

\textbf{Description: }In the process of software evolution, due to the increasing functional requirements of the system and the poor programming practices of developers, some classes may not be well reconstructed and optimized, making them very bloated. Thus, the \textit{LILBC} occurs.

\textbf{Impact: }The concept of a \textit{God Class} has been extensively studied, while the \textit{LILBC} presents a distinct challenge as it involves interactions between multiple PLs. In the context of extending Python with C++ using pybind11, it is a common practice to bind C++ classes to Python modules using the \textsc{class\_} template class and then use \textsc{def} to bind member functions of the class. This process of binding member functions to the class can be divided into two types. One is to bind directly behind the template class. The code of this type of binding appears in the same location and the binding process is clear. The other is to use aliases for indirect binding, which can be used anywhere for this purpose. Adding member functions to a class provides greater flexibility. No matter which type of binding is used, when a class is bound to too many member functions, it will make the class difficult to understand and maintain. Large classes usually have strong dependencies on other classes, which leads to increased coupling between different classes, making code modifications more difficult and complex.

\textbf{Refactoring Recommendation: }Split the large C++ class into multiple classes with clear responsibilities and bind them to Python modules respectively.

\textbf{Identification Method: }
\begin{equation}\tag{Rule6}\label{eq6}
\footnotesize{
\begin{gathered}
NBFunctions(c)+NBFunctions(alias(c))>\\ \hfill MaxNBFunctionsThreshold
\end{gathered}
}
\end{equation}

The native class $c$ is defined in~\ref{eq1}. When the total number of directly bound functions in a class and the number of functions indirectly bound by the class using an alias is greater than a predefined threshold for a class, the \textit{LILBC} is introduced. It is worth noting that there are many ways to bind functions using pybind11, such as \textsc{def}, \textsc{def\_readonly}, \textsc{def\_property}, etc. When counting functions bound in classes, we exclude built-in functions (e.g., Listing~\ref{lst:code7_1}, the C++ class \textsc{RowTensor} binds the two built-in functions \textsc{\_\_str\_\_} and \textsc{\_\_repr\_\_} to the corresponding Python module~\citep{mindspore_lilbc}), which in Python are identified by double underscores at both the beginning and end of their names. This is because these functions are called by the Python parser in specific cases, rather than being called directly by the code written by the developer. Even if these functions are not defined, Python will provide default implementations. We set $MaxNBFunctionsThreshold$ as 7, following the work by Abidi \textit{et al}~\citep{abidi2021multi}, and 7 was regarded as the number of objects that the human brain can understand~\citep{lippert2006refactoring}.

\begin{lstlisting}[language=C++,mathescape=true,breaklines=true,firstnumber=1]
(void)py::class_<RowTensor, std::shared_ptr<RowTensor>>(*m, "RowTensor")
\end{lstlisting}
{\lstset{aboveskip=0pt,belowskip=0pt}
\begin{lstlisting}[language=C++,numbers=none]
...
\end{lstlisting}
}
\begin{lstlisting}[language=C++,caption=Bind built-in functions,label={lst:code7_1},mathescape=true,breaklines=true,firstnumber=13]
    .def("__str__", &RowTensor::ToString)
    .def("__repr__", &RowTensor::ToString);   
\end{lstlisting}

\textbf{Example: }
\textsc{PyClient} is a C++ class bound using pybind11, which is called in Python through the function name \textsc{Client}, as shown in Listing~\ref{lst:code7}. Twenty-nine functions are bound to this class, and the binding process employs the alias \textsc{py\_local\_client}~\citep{tensorflow_lilbc}. Despite this, we have not observed bindings through this alias elsewhere.
\begin{lstlisting}[language=C++,mathescape=true,breaklines=true,firstnumber=1]
py::class_<PyClient, std::shared_ptr<PyClient>> py_local_client(m, "Client");
  py_local_client.def_property_readonly("platform", &PyClient::platform_name)
\end{lstlisting}
{\lstset{aboveskip=0pt,belowskip=0pt}
\begin{lstlisting}[language=C++,numbers=none]
...
\end{lstlisting}
}
\begin{lstlisting}[language=C++,caption=Design Smell - Large Inter-language Binding Class,label={lst:code7},mathescape=true,breaklines=true,firstnumber=74]
      .def("emit_python_callback", &PyClient::EmitPythonCallback,
           py::arg("callable"), py::arg("builder"), py::arg("operands"),
           py::arg("result_shapes"), py::arg("operand_layouts") = std::nullopt,
           py::arg("has_side_effects") = false);
\end{lstlisting}

\subsection{Excessive Inter-Language Communication (EILC)}

\textbf{Description: }In Python files, it is common to call C/C++ entities to achieve high-performance computing and implement performance optimizations. Such inter-language calls are considered as a form of language communication between Python and C/C++. However, as the functional requirements of a Python file expand, developers may often resort to simply adding calls to the relevant C/C++ entities without considering the frequency of these calls.

\textbf{Impact: }When a Python file excessively calls native C/C++ entities without proper optimization, it can lead to increased coupling between the Python file and multiple C/C++ files. When a bug occurs in the Python file, it will be difficult to promptly identify and debug, thus increasing the workload for troubleshooting and resolution. Furthermore, due to the high level of coupling, modifying the Python file may necessitate changes to multiple C/C++ files, seriously affecting code maintainability.

\textbf{Refactoring Recommendation: }Split the Python code into multiple modules, each of which is responsible for independent functions and tasks, which makes the code structure clearer and easier to manage. 

\textbf{Identification Method: }
\begin{equation}\tag{Rule7}\label{eq7}
\footnotesize{
\begin{gathered}
\alpha:=IsCalled(f,alias(f),c,alias(c))=True \\
Unique(FileCalls(\alpha))>MaxNBCallsFilesThreshold
\end{gathered}
}
\end{equation}
We define $\alpha$ to indicate that a native function $f$ or a native class $c$ is called, including direct and indirect calls. $f$ and $c$ are defined in~\ref{eq1}.
This ILDS occurs when, in a Python file, the total number of distinct native files associated with $\alpha$ exceeds the maximum threshold of all natively referenced files in an external file's call. 

\begin{figure}[h]
  \centering
  \includegraphics[width=\linewidth]{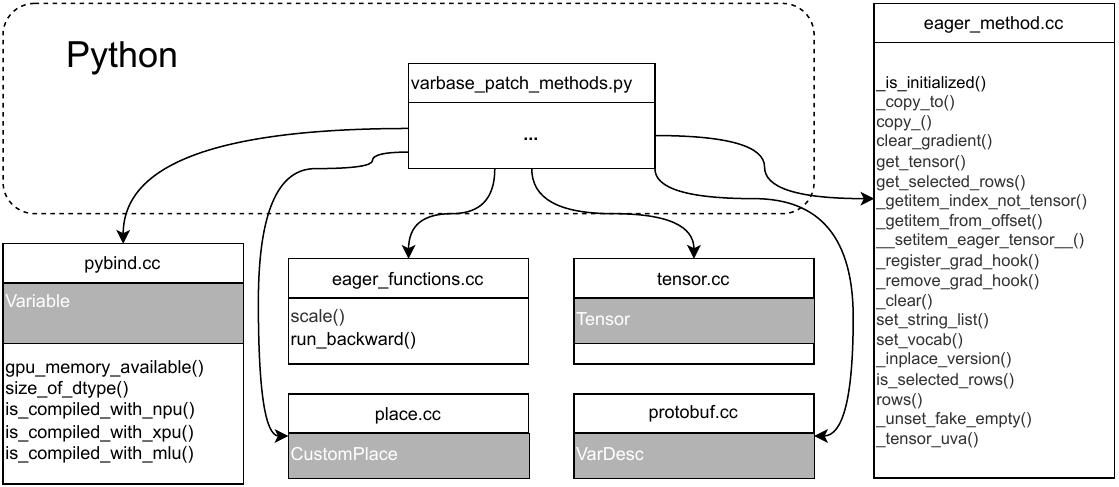}
  \caption{An example of design smell \textit{Excessive Inter-Language Communication}}\label{fig:f3}
\end{figure}

\textbf{Example: }\textsc{varbase\_patch\_methods.py} is a Python file in PaddlePaddle 2.4.2 that processes dynamic graphics~\citep{paddlepaddle_eilc}. As shown in Figure.~\ref{fig:f3}, this Python file communicates with 6 native C++ files, including calling 26 native functions and 4 native classes, which results in excessive coupling.

\section{Implementation and Validation}\label{sec_tool}

\subsection{Implementation of ILDS Detection}
\begin{figure*}[h]
  \centering
  \includegraphics[width=0.7\textwidth]{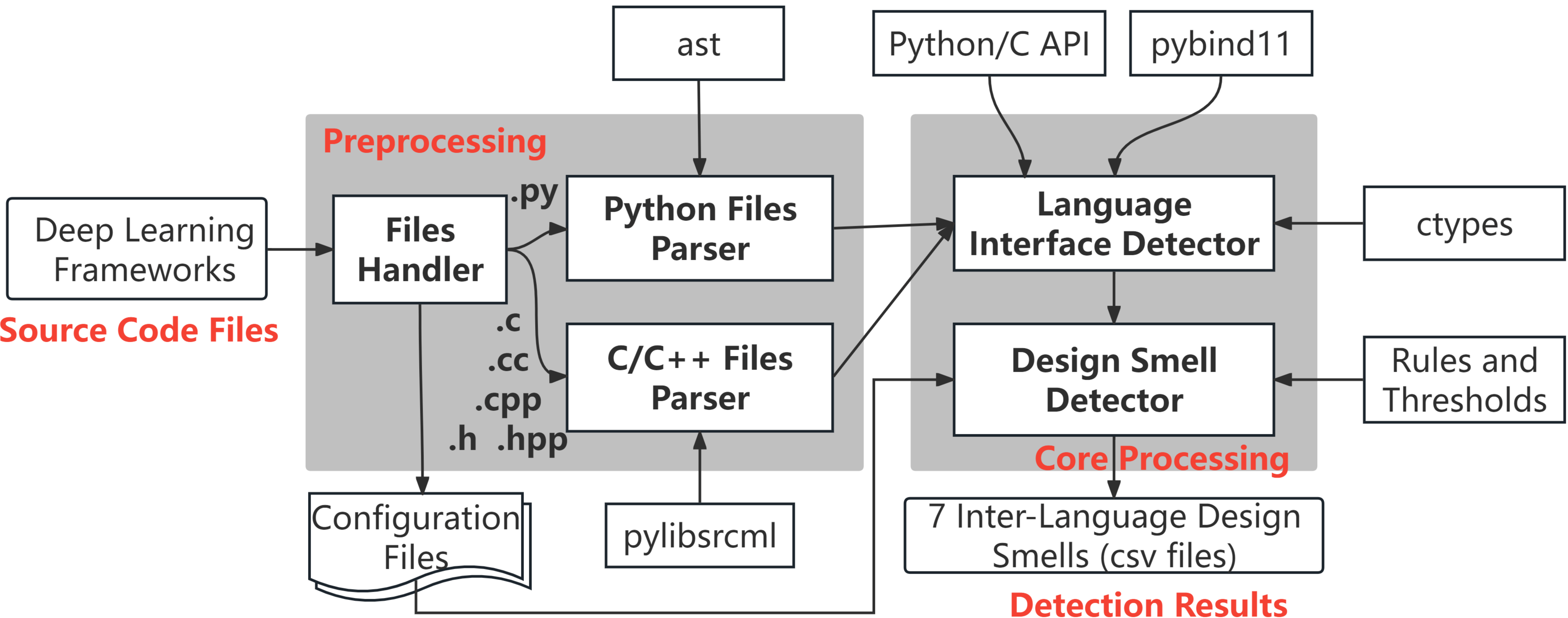}
  \caption{Execution process of the \textsc{CPsmell} tool}
  \label{fig:f4}
\end{figure*}

To implement the detection rules for the 7 ILDS defined in Section \ref{sec_ILDS}, we developed the \textsc{CPsmell} toolkit~\citep{cpsmell}. The entire execution process of \textsc{CPsmell} is shown in Figure~\ref{fig:f4}. The main processing part includes four steps. First, it gets as input the source files of a multi-language software system written in Python and C/C++, such as a DLF. The source files are processed and divided into Python files and C/C++ files according to the file extension name. Second, C/C++ files are parsed into XML files by the pylibscrml toolkit for archiving analysis. \textsc{pylibsrcml}~\citep{pylibsrcml} is the binding of \textsc{libsrcml} in Python, supporting all \textsc{libsrcml} functionalities. \textsc{libsrcml} can convert source code files into XML representation, supporting multiple PLs, such as C, C++, and Java. In addition, Python files are parsed into Abstract Syntax Trees (ASTs)~\citep{ast} using the Python built-in module \textsc{ast}. Third, the language interface detector is compatible with the detection of three ILCMs: Python/C API, pybind11, and ctypes. 
When the language interface detector detects Python/C API and pybind11, it captures the C++ extension modules as the bridge between Python and C/C++ and detects all classes and functions defined in these extension modules for interaction. The implementation specifically involves custom AST and the use of XPath for addressing within XML files. ctypes is relatively rarely used in these DLFs and is typically used to call C/C++ functions from DLLs. The language interface detector analyzed all Python files containing \textsc{import ctypes} using regular expressions. Finally, we designed a design smell detector to analyze the language interface between Python and C/C++ based on the rules and thresholds that we defined. 

\textsc{CPsmell} will save in CSV files all intermediate detection results, such as names, paths, and other related information of modules, classes, and functions that interact between Python and C/C++. In addition, \textsc{CPsmell} explicitly records information about each identified ILDS instance, facilitating developers in locating the specific files and code lines associated with each smell.

\vspace*{-3pt}
\subsection{Validation Design}

To validate the effectiveness of the \textsc{CPsmell} tool, we asked the following research question: \textbf{\textit{How accurate is \textsc{CPsmell} in detecting inter-language design smells?}}

\subsubsection{Data Collection.}
We chose DLFs as the dataset for two main reasons. 
First, research on the software quality issues of DLFs is still lacking, despite their rapid evolution in response to the growing demand for model and algorithm optimization due to the rise of artificial intelligence in various fields.
Second, most DLFs use multi-language programming with Python and C/C++, which are representative of multi-language systems involving Python and C/C++.

We then followed two inclusion criteria to select each DLF: firstly, the DLF must use at least two PLs, with Python and C/C++ constituting no less than 20\% of the codebase; secondly, the DLF uses at least two of the following three ILCMs: Python/C API, pybind11, and ctypes. Ultimately, we identified 5 popular DLFs, namely TensorFlow~\citep{abadi2016tensorflow}, PyTorch~\citep{paszke2019pytorch}, Chainer~\citep{tokui2015chainer}, PaddlePaddle~\citep{ma2019paddlepaddle}, and MindSpore~\citep{huawei2022huawei}. 
We collected the latest release version for each selected DLF for this tool validation, i.e., TensorFlow 2.12.0, PyTorch 2.0.1, Chainer 7.8.1, PaddlePaddle 2.4.2, and MindSpore 1.10.1.
Table \ref{tab:t1} shows the proportion of primary PLs used and the ILCMs adopted in the 5 DLFs. Then, we used \textsc{CPsmell} to detect ILDS instances in the collected versions of the 5 DLFs.

\begin{table*}[htbp]
\caption{Proportion of PLs used and ILCMs adopted in DLFs}
\label{tab:t1}
\scalebox{1.0}{
\begin{tabular}{@{}ccccccc@{}}
\toprule
 & \textbf{\%Python} & \textbf{\%C/C++} & \textbf{\%Total} & \textbf{Python/C API} & \textbf{pybind11} & \textbf{ctypes}\\ \midrule
TensorFlow   & 20.8              & 63.1             & 83.9            & \checkmark  & \checkmark        & \checkmark \\
PyTorch      & 45.5              & 46.9             & 92.4            & \checkmark  & \checkmark        & \checkmark \\
Chainer      & 76.3              & 20.0             & 96.3            &              & \checkmark        & \checkmark \\
PaddlePaddle & 44.1              & 46.4             & 90.5            & \checkmark  & \checkmark        & \checkmark \\
MindSpore    & 28.0              & 66.1             & 94.1            &              & \checkmark        & \checkmark \\ \bottomrule
\end{tabular}
}
\end{table*}


\subsubsection{Data Analysis.}\label{sec_dataanalysis_IV}
To alleviate personal biases, two authors participated in the ILDS validation, and both authors had relevant experience in inter-language programming. The specific verification steps are described as follows:

\noindent \textbf{Step1: Pilot labeling.} We randomly sampled from all ILDS instances by selecting a quantity (greater than 0 but not exceeding 50\%) of instances for each ILDS. In total, 100 ILDS instances were randomly selected for validation, and each validation result was labeled as either the identified ILDS or not. The two authors validated the 100 ILDS instances independently and obtained two sets of validation results.

\noindent \textbf{Step2: Calculating the Kappa value.} We compared the two sets of validation results and calculated the Kappa value as a measure of result similarity~\citep{viera2005understanding}. If the Kappa value >= 0.8, it was considered a consensus between the two authors. Any disagreements were discussed to reach a consensus on the 100 samples. If the Kappa value < 0.8, we first discussed the discordant samples to eliminate discrepancies and then repeated steps 1 and 2 with the remaining samples until the Kappa value reached or exceeded 0.8.

\noindent \textbf{Step3: Labeling the remaining data.} Divide the remaining samples into two groups, with two authors individually labeling one group of samples each. Finally, the labeling results were consolidated and the accuracy was calculated.

The entire validation process mainly adheres to two criteria. First, a true ILDS instance should fully conform to the detection rule defined in Section \ref{sec_ILDS} for the corresponding ILDS. Second, combining the code context to determine if it falls into a special case, i.e., situations not covered by the rules. 

We calculated the accuracy of each ILDS in the 5 DLFs. 
$TP$ represents the number of ILDS instances detected by \textsc{CPsmell} that we considered correct, $FP$ represents the number of ILDS instances detected by \textsc{CPsmell} that we considered incorrect, and the calculation formula is: $Accuracy=TP/(TP+FP)$.

\subsection{Validation Results}
Using \textsc{CPsmell}, 982 ILDS instances were detected in the latest releases of the 5 DLFs. 
After the first round of pilot labelling, we got the Kappa value of 0.82, which means the two authors reached a consensus on the labelling of true ILDS instances. The final labelling results are shown in Table \ref{tab:t10}. 
NA represents the case of $TP+FP=0$, that is, the number of ILDS instances detected by \textsc{CPsmell} is 0. 

\begin{table*}[htbp]
\caption{Detection accuracy of \textsc{CPsmell}}
\label{tab:t10}
\scalebox{1.0}{
\begin{tabular}{@{}crrrrrrrr@{}}
\toprule
              & \textbf{LLF}      & \textbf{UNE}     & \textbf{LREC}     & \textbf{LSD}      & \textbf{NURP}    & \textbf{LILBC}    & \textbf{EILC}     & \textbf{ALL}     \\ \midrule
TP           & 488      & 236     & 9        & 127      & 15      & 70       & 19       & 964     \\
TP+FP         & 488      & 253     & 9        & 127      & 16      & 70       & 19       & 982     \\
Accuracy  & 100\% & 93.28\% & 100\% & 100\% & 93.75\% & 100\% & 100\% & 98.17\% \\\bottomrule
\end{tabular}
}
\end{table*}

When taking all ILDS of all DLFs as a whole, the overall accuracy of ILDS detection by \textsc{CPsmell} is 98.17\%, which is a very high level of accuracy. 
From the perspective of DLFs, the accuracy of detection for all ILDS is greater than 96\% for each DLF.
From the perspective of ILDS, 
except for \textit{UNE} (93.28\%) and \textit{NURP} (93.75\%), the accuracy of detection for other six ILDS reaches 100\%. We published our validation data online\footnote{https://github.com/ILDSOFDLF/cpsmell/tree/master/validation}.

\textbf{Summary: }Our validation shows that \textsc{CPsmell} performs excellently in detecting ILDS, with an overall accuracy of 98.17\%.


\subsection{Validation Results Analysis} \label{sec_IV_diss}
In general, \textsc{CPsmell} demonstrates a commendable level of ILDS detection accuracy. However, it made a mistake when detecting \textit{NURP}. Through analysis, we observed that developers use absolute or non-relative paths for loading DLLs in some cases. Yet, they typically add conditional checks in the code's context to avoid potential errors. These situations will not be included. For example, in Listing~\ref{lst:code-mistake1}, the developers used the absolute path to load the DLL, but before doing so, they used the \textsc{os.path.exists()} function to determine whether the path exists~\citep{pytorch_discuss1}. 

\begin{lstlisting}[language=Python,caption=Misjudgment - Not Using Relative Path,label={lst:code-mistake1},mathescape=true,breaklines=true]
if "gomp" in str(e) and os.path.exists("/usr/lib64/libgomp.so.1"):
# hacky workaround for fbcode/buck
  global _libgomp
  libgomp = cdll.LoadLibrary("/usr/lib64/libgomp.so.1")
\end{lstlisting}

The detection inaccuracy of \textsc{CPsmell} primarily stems from its \textit{UNE} detection. In PyTorch, for instance, we observed instances of false positives due to developers using functions similar to \textsc{hasattr()}, as shown in Listing~\ref{lst:code-mistake2}, three native functions are stored as strings in a tuple, and \textsc{hasattr()} is used to determine their association with the \textsc{torch.\_C} module, which is a C/C++ extension module~\citep{pytorch_discuss2}. These functions appear in the form of strings, so \textsc{CPsmell} is difficult to accurately identify. Additionally, since we cannot determine the total number of ILDS in DLFs, we only examined the accuracy of the tool and not the recall. However, our tool achieves the identification of three commonly used ILCMs, which we believe can cover most situations.

\begin{lstlisting}[language=Python,caption=Misjudgment - Unused Native Entity,label={lst:code-mistake2},mathescape=true,breaklines=true]
valgrind_symbols = (
    "_valgrind_supported_platform",
    "_valgrind_toggle",
    "_valgrind_toggle_and_dump_stats",
)
if all(hasattr(torch._C, symbol) for symbol in valgrind_symbols):
    self._supported_platform: bool = torch._C._valgrind_supported_platform()
\end{lstlisting}

\vspace*{-3pt}
\section{Empirical Study}\label{sec_empiricalstudy}
The objective of this empirical study is to comprehensively evaluate the 5 multi-language DLFs in terms of ILDS, thereby having a deep understanding on the state of the maintainability of the DLFs.

\subsection{Study Design}
\subsubsection{Research Questions}\label{sec_RQ}
To explore the distribution of ILDS in the DLFs, and to understand the developers' awareness of ILDS as well as the reasons for the introduction of ILDS, we formulate the following three research questions (RQs). 

\noindent\textbf{RQ1: What is the distribution of different inter-language design smells in DLFs?}
By calculating the distribution of each ILDS, we can gain a clear understanding of the severity of different design smells in DLFs. This, in turn, can enhance developers' awareness and attention to design smells that are more widely distributed.

\noindent\textbf{RQ2: What is the proportion of inter-language design smells that are fixed in DLFs?}
Studying the fix ratio of ILDS in DLFs can get informed whether certain smells have been fixed as DLFs evolve, thus indicating the extent to which developers have progressively become aware of these smells.

\noindent\textbf{RQ3: How do inter-language design smells evolve in DLFs?}
While we already know whether some smells have been fixed, many of them are introduced at later stages. Studying the evolution of ILDS in DLFs can provide us with a clear understanding of the introduction process for each smell, and enable further analysis of the reasons behind their introduction.

\subsubsection{Data Collection}
To address the RQs formulated in Section \ref{sec_RQ}, we collected ILDS instances and related data from the historical versions of the 5 DLFs on GitHub using \textsc{CPsmell}. In the early versions, most of these DLFs employed customized inter-language communication methods, but these methods were immature, while \textsc{CPsmell} was based on the three relatively mature ILCMs, resulting in fewer detected ILDS instances. Therefore, we selected the latest versions released between 2020 and 2023. In each DLF, we selected 10 versions, and the time interval between each two adjacent versions is about 3 months. The specific version information is shown in Table \ref{tab:t8}, where $V_s$ and $T_s$ represent the version and the release date of the start version respectively, and $V_e$ and $T_e$ represent the version and the release date of the end version respectively. In particular, the latest version of Chainer was released in 2022, thus we collected Chainer’s data from 2019 to 2022. 

\begin{table*}[htbp]
\caption{Version information of DLFs for data collection}
\label{tab:t8}
\scalebox{1.0}{
\begin{tabular}{@{}cccccc@{}}
\toprule
 & \textbf{TensorFlow} & \textbf{PyTorch} & \textbf{Chainer} & \textbf{PaddlePaddle} & \textbf{MindSpore} \\ \midrule
$V_s$            & 2.2.0               & 1.5.0            & 6.0.0            & 1.8.0           & 1.0.0              \\ 
$T_s$             & 2020-05-06          & 2020-04-22       & 2019-05-16       & 2020-04-30      & 2020-09-23         \\ 
$V_e$            & 2.12.0              & 2.0.1            & 7.8.1            & 2.4.2           & 1.10.1             \\ 
$T_e$            & 2023-03-23          & 2023-05-09       & 2022-01-05       & 2023-02-15      & 2023-02-16         \\ \bottomrule
\end{tabular}
}
\end{table*}

\vspace*{-3pt}
\subsubsection{Data Analysis}\label{sec_dataanalysis_ES}
For RQ1, we analyzed each DLF from three aspects: the number and proportion of ILDS instances, the number and proportion of smelly files, and the density of ILDS instances in files and code lines. The specific calculation data are shown in Table \ref{tab:t3}.
For RQ2, we used formula $r_{i,j}=1-\vert{S_i\cap S_j}\vert/\vert{S_i}\vert$ 
to represent the proportion of ILDS fixes, where $r_{i,j}$ denotes the rate of ILDS fixes from version $i$ to version $j$, and $S_i$, $S_j$ denote the sets of ILDS instances in versions $i$, $j$, respectively.
For RQ3, we counted the number of different ILDS instances in each DLF, considering 10 versions for each DLF. This allowed us to study the trends of ILDS across different versions.

\begin{table*}[htbp]
\caption{Data items to be calculated (RQ1)}
\label{tab:t3}
\scalebox{1.0}{
\begin{tabular}{clc}
\toprule
\textbf{Data item} & \multicolumn{1}{c}{\textbf{Description}}                                                                                                                                                 & \textbf{Related table} \\ \midrule
$f$         & \begin{tabular}[c]{@{}l@{}}The number of files involving inter-language \\ communication. The files specifically refer to those \\defining C/C++ entities, files in Python that call \\ these entities, and files using ctypes.\end{tabular}                                                                                                         & Table \ref{tab:t4}               \\ 
$l$         & \begin{tabular}[c]{@{}l@{}}The number of code lines involving inter-language \\ communication, i.e., the total number of code lines \\ contained in $f$.\end{tabular} & Table \ref{tab:t4}               \\ 
$s_i$         & The number of instances of ILDS $i$.                                                                                                                                 & Table \ref{tab:t5}               \\ 
$\frac{s_i}{\sum_{i=1}^{7}s_i}$         & The proportion of instances of ILDS $i$.                                                                                                                             & Table \ref{tab:t5}               \\ 
$\widetilde{f}_i$         & Number of files containing  instances of ILDS $i$.                                                                                                                                & Table \ref{tab:t6}               \\ 
$\frac{\widetilde{f}_i}{f}$         & Proportion of files containing  instances of ILDS $i$.                                                                                                                            & Table \ref{tab:t6}               \\ 
$unique(\sum_{i=1}^{7}\widetilde{f}_i)$         & \begin{tabular}[c]{@{}l@{}}Number of all files containing ILDS instances,\\ where $unique()$ removes duplicates.\end{tabular}                                                 & Table \ref{tab:t6}               \\ 
$\frac{unique(\sum_{i=1}^{7}\widetilde{f}_i)}{f}$         & Proportion of all files containing ILDS instances.                                                                                                                           & Table \ref{tab:t6}               \\ 
$\frac{\sum_{i=1}^{7}s_i}{f}$        & Density of ILDS instances in files.                                                                                                                                         & Table \ref{tab:t7}               \\ 
$\frac{\sum_{i=1}^{7}s_i}{l}$        & Density of ILDS instances in code lines.                                                                                                                                    & Table \ref{tab:t7}               \\ 
\bottomrule
\end{tabular}
}
\end{table*}

\vspace*{-3pt}
\subsection{Results }\label{s:5}

 \subsubsection{Distribution of Different Inter-Language Design Smells in DLFs (RQ1)}
We selected recent and stable versions of the DLFs and used \textsc{CPsmell} to detect the number of files and code lines involving inter-language communication in the 5 DLFs, as shown in Table \ref{tab:t4}. We answer RQ1 in the following three aspects.

\textbf{(1) The number and proportion of instances of different ILDS.} As shown in Table \ref{tab:t5}, among the 5 DLFs, TensorFlow, PyTorch and PaddlePaddle all have more than 100 ILDS instances. And TensorFlow is detected with the largest number of ILDS instances, i.e., 480 instances. \textit{LSD} is more prominent in PyTorch, accounting for 41.20\% of all ILDS. MindSpore exhibits a higher prevalence of \textit{LILBC}, accounting for 73.91\%. In addition, \textit{LLF} accounts for the highest proportion in the 5  DLFs.

\begin{table*}[htbp]
\caption{Number of files and code lines involved in inter-language communication (RQ1)}
\label{tab:t4}
\scalebox{1.0}{
\begin{tabular}{@{}cccccc@{}}
\toprule
        & \textbf{TensorFlow} & \textbf{PyTorch} & \textbf{Chainer} & \textbf{PaddlePaddle} & \textbf{MindSpore} \\ \midrule
Number Of Files      & 287                 & 187              & 23               & 185                   & 166                \\
Number Of Code Lines & 118,358              & 96,171            & 3,232             & 63,928                 & 130,971             \\ \bottomrule
\end{tabular}
}
\end{table*}

\begin{table*}[htbp]
\caption{Number and proportion of ILDS instances (RQ1)}
\label{tab:t5}
\scalebox{1.0}{
\begin{tabular}{@{}crrrrrrrrrrrr@{}}
\toprule
 & \multicolumn{2}{r}{\textbf{TensorFlow}} & \multicolumn{2}{r}{\textbf{PyTorch}} & \multicolumn{2}{r}{\textbf{Chainer}} & \multicolumn{2}{r}{\textbf{PaddlePaddle}} & \multicolumn{2}{r}{\textbf{MindSpore}} & \multicolumn{2}{r}{\textbf{All}} \\ 
\textbf{ILDS} & \textbf{\#}        & \textbf{\%}        & \textbf{\#}       & \textbf{\%}      & \textbf{\#}       & \textbf{\%}      & \textbf{\#}         & \textbf{\%}         & \textbf{\#}        & \textbf{\%}       & \textbf{\#}     & \textbf{\%}               \\ \midrule
UNE          & 137                 & 28.04                & 48               & 15.94               & 6               & 40.00                  & 57                  & 34.97                    & 5                & 21.74                   & 253             & 25.76             \\
LLF          & 325               & 67.71               & 93              & 30.90              & 9               & 60.00              & 61                 & 37.42                & 0                & 0                   & 488           & 49.69            \\
LREC         & 0                 & 0                   & 9               & 3.00               & 0               & 0                  & 0                  & 0                    & 0                & 0                   & 9             & 0.92             \\
LSD          & 1                 & 0.21                & 124             & 41.20              & 0               & 0                  & 2                  & 1.23                 & 0                & 0                   & 127           & 12.93            \\
NURP         & 3                 & 0.63                & 11              & 3.70               & 0               & 0                  & 1                  & 0.61                 & 1                & 4.35                & 16            & 1.63             \\
LILBC        & 11                & 2.29                & 1               & 0.33               & 0               & 0                  & 41                 & 25.15                & 17               & 73.91               & 70            & 7.13             \\
EILC         & 3                 & 0.63                & 15              & 4.98               & 0               & 0                  & 1                  & 0.61                 & 0                & 0                   & 19            & 1.93             \\
ALL          & 480               & 100.00              & 301             & 100.00             & 15              & 100.00             & 163                & 100.00               & 23               & 100.00              & 982           & 100.00           \\ \bottomrule
\end{tabular}
}
\end{table*}

\textbf{(2) The number and proportion of smelly files for different ILDS.} As shown in Table \ref{tab:t6}, there is significant variation in the distribution of smells across different DLFs. The proportion of smelly files of TensorFlow, PyTorch, and Chainer exceeds 27\%, while MindSpore has a much lower proportion of smelly files.

\begin{table*}[htbp]
\caption{Number and proportion of ILDS files (RQ1)}
\label{tab:t6}
\scalebox{1.0}{
\begin{tabular}{@{}crrrrrrrrrr@{}}
\toprule
 & \multicolumn{2}{r}{\textbf{TensorFlow}} & \multicolumn{2}{r}{\textbf{PyTorch}} & \multicolumn{2}{r}{\textbf{Chainer}} & \multicolumn{2}{r}{\textbf{PaddlePaddle}} & \multicolumn{2}{r}{\textbf{MindSpore}} \\ 
\textbf{ILDS }        & \textbf{\#}                & \textbf{\%}                  & \textbf{\#}              & \textbf{\%}                 & \textbf{\#}              & \textbf{\%}                 & \textbf{\#}                 & \textbf{\%}                   & \textbf{\#}                & \textbf{\%}                 \\\midrule
UNE          & 34                 & 11.84                & 15               & 8.02               & 4               & 17.39                  & 17                  & 9.19                    & 5                 & 3.01                  \\
LLF          & 52                & 18.12               & 5               & 2.67               & 6               & 26.09              & 8                  & 4.32                 & 0                 & 0                  \\
LREC         & 0                 & 0                   & 7               & 3.74               & 0               & 0                  & 0                  & 0                    & 0                 & 0                  \\
LSD          & 1                 & 0.35                & 17              & 9.09               & 0               & 0                  & 1                  & 0.54                 & 0                 & 0                  \\
NURP         & 1                 & 0.35                & 7               & 3.74               & 0               & 0                  & 1                  & 0.54                 & 1                 & 0.6                \\
LILBC        & 7                 & 2.44                & 1               & 0.53               & 0               & 0                  & 15                 & 8.11                 & 10                & 6.02               \\
EILC         & 16                & 5.57                & 37              & 19.79              & 0               & 0                  & 6                  & 3.24                 & 0                 & 0                  \\
ALL          & 78                & 27.18               & 65              & 34.76              & 7               & 30.43              & 31                 & 17.3                 & 13                & 7.83               \\ \bottomrule
\end{tabular}
}
\end{table*}

\textbf{(3) ILDS density in files and lines of code.} Due to substantial differences in the scale of different DLFs, we conducted statistics at the two granularities of files and code lines to better illustrate the overall distribution of ILDS instances. As shown in Table \ref{tab:t7}, TensorFlow and PyTorch exhibit the highest ILDS density in files, Chainer has the highest ILDS density in code lines, while MindSpore shows a lower ILDS density in both files and code lines. In addition, the ILDS density in code lines of PaddlePaddle (ranked 4th) is 14.7 times higher than that of MindSpore (ranked 5th).

\begin{table*}[htbp]
\caption{ILDS Density in code lines and files (RQ1)}
\label{tab:t7}
\scalebox{1.0}{
\begin{tabular}{@{}cccccc@{}}
\toprule
 & \textbf{TensorFlow} & \textbf{PyTorch} & \textbf{Chainer} & \textbf{PaddlePaddle} & \textbf{MindSpore} \\ \midrule
File Density & 1.67                & 1.61             & 0.65             & 0.88                  & 0.14               \\
KLOC Density & 4.06                & 3.13             & 4.64             & 2.55                  & 0.18               \\ \bottomrule
\end{tabular}
}
\end{table*}

\begin{framed}
\noindent\textbf{Answer to RQ1: }There are significant differences in the distributions of various ILDS between the 5 DLFs, with \textit{LLF} and \textit{UNE} being the most prevalent ILDS in DLFs, each ILDS accounting for over 25\% of all the instances in each DLF. Certain ILDS occur frequently in a particular DLF, such as 41.2\% of the ILDS instances in PyTorch is \textit{LSD}, while the proportion of this ILDS in other DLFs does not exceed 1.5\%.
\end{framed}

\subsubsection{Proportion of Fixed Inter-Language Design Smells Instances in DLFs (RQ2)}

The proportion of ILDS fixes is shown in Table \ref{tab:t9}.
It is evident that many data cells in Table \ref{tab:t9} are marked as NA, indicating that these ILDS were not present in the version released three years ago. As the DLF continued to evolve, the size of the file package increased, leading to the introduction of new ILDS instances. Additionally, the data in Table \ref{tab:t9} show significant variations in the proportion of ILDS fixes among different DLFs for ILDS instances that existed three years ago. Overall, \textit{EILC}, \textit{LILBC}, and \textit{UNE} are with the highest average fix ratios. It is worth noting that \textit{EILC}’s fix rate reaches 100\%, which is attributed to PyTorch. Through analyzing the releases of PyTorch over the past three years, we found significant architectural changes, making \textit{EILC} (also as an architectural smell) more noticeable and likely to be resolved. However, \textit{LREC}, and \textit{NURP} have not received effective fixes over the three-year version iterations (all 0\%). In addition, among the 5 DLFs, PaddlePaddle has the highest overall ILDS fix rate, reaching 18.52\%, while the overall ILDS fix rates of TensorFlow and PyTorch are below 10.00\%.

\begin{framed} 
\noindent\textbf{Answer to RQ2: }We found that some ILDS are more likely to be fixed, such as \textit{EILC}, \textit{LILBC}, and \textit{UNE}. While some smells are more prone to being overlooked, such as \textit{LREC}, and \textit{NURP}.
\end{framed}

\begin{table*}[htbp]
\caption{Proportion of fixed ILDS instances (RQ2)}
\label{tab:t9}
\scalebox{1.0}{
\begin{tabular}{@{}crrrrrr@{}}
\toprule
 & \textbf{TensorFlow} & \textbf{PyTorch} & \textbf{Chainer} & \textbf{PaddlePaddle} & \textbf{MindSpore} & \textbf{ALL}                                    \\ \midrule
UNE         & 12/66, 18.18\%                 & 5/41, 12.20\%              & 0/7, 0\%               & 4/15, 26.67\%         & 0/2, 0\%                 & 21/131, 16.03\%                                             \\ 
LLF         & 6/193, 3.11\%              & 1/73, 1.37\%           & 2/9, 22.22\%          & 5/22, 22.73\%         & NA                 & 14/297, 4.71\%                                          \\ 
LREC        & NA                  & 0/6, 0\%              & NA               & NA              & NA                 & 0/6, 0\%                                             \\ 
LSD         & 0/1, 0\%                 & 4/61, 6.56\%           & NA               & NA              & NA                 & 4/62, 6.45\%                                          \\ 
NURP        & 0/1, 0\%                 & 0/1, 0\%              & NA               & NA              & NA                 & 0/2, 0\%                                             \\ 
LILBC       & 3/5, 60.00\%             & 0/1, 0\%              & 1/1, 100.00\%         & 1/17, 5.88\%          & 1/5, 20.00\%            & 6/29, 20.69\%                                         \\ 
EILC        & NA                  & 2/2, 100.00\%            & NA               & NA              & NA                 & 2/2, 100.00\%                                        \\ 
ALL         & 21/266, 7.89\%              & 12/185, 6.49\%           & 3/17, 17.65\%          & 10/54, 18.52\%         & 1/7, 14.29\%            &47/519, 7.90\%  \\ \bottomrule
\end{tabular}
}
\end{table*}

\subsubsection{Evolution of Inter-Language Design Smells in DLFs (RQ3)}
\begin{figure*}[h]
  \centering
  \includegraphics[width=0.7\textwidth]{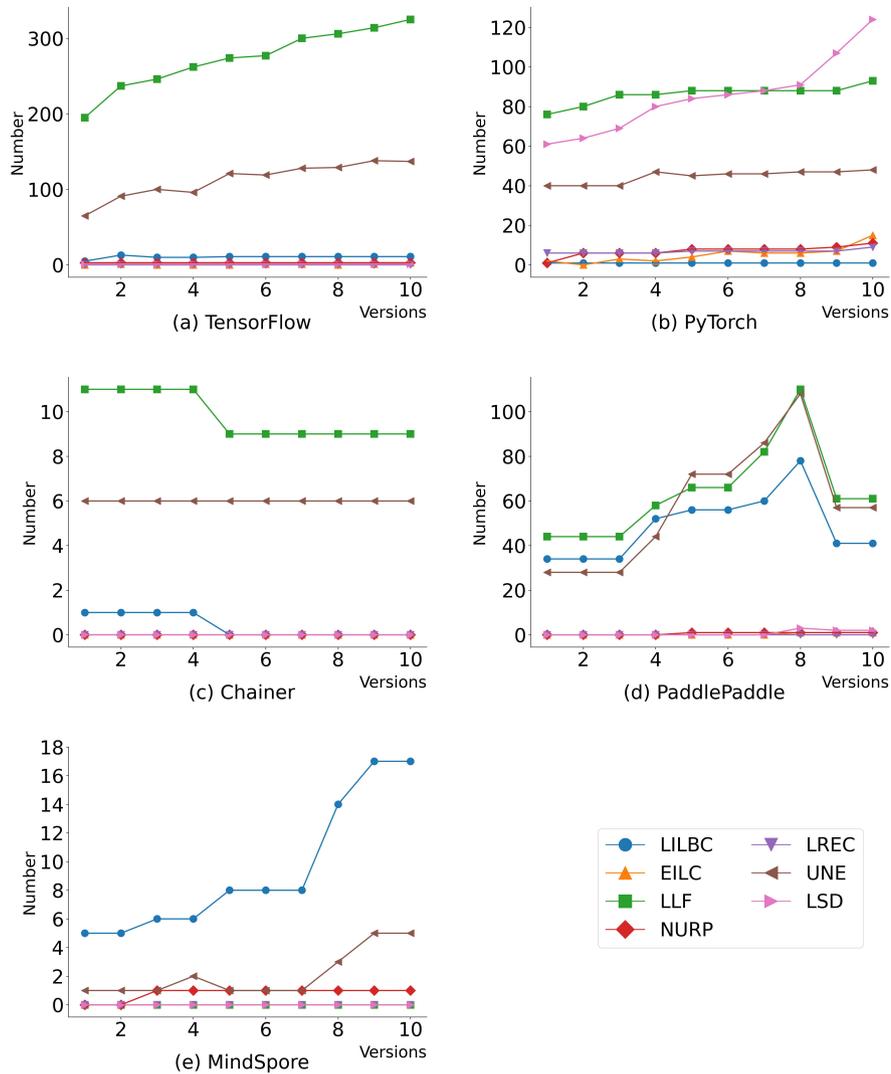}
  \caption{The evolution of ILDS in the 5 DLFs (RQ3)}\label{fig:f5}
\end{figure*}

To further clarify the changes of ILDS, we investigated the evolution of ILDS in RQ3. 
We employed \textsc{CPsmell} to analyze the source code of 10 versions from each of the 5 DLFs. We further investigated the trends in the number of various ILDS instances as the DLFs evolved, and the results are depicted in Figure \ref{fig:f5}. The time range represented by the 10 versions on the horizontal axis is shown in Table \ref{tab:t8}, with a time interval of 3 months between adjacent versions. The vertical axis in Figure \ref{fig:f5} shows the number of ILDS instances. 
For Chainer, except for the decrease of \textit{LLF} and \textit{LILBC} instances from the 4th to 5th version, the overall number of ILDS instances remains relatively stable, with minor variations. In TensorFlow, PyTorch, PaddlePaddle, and Mindspore, the number of ILDS instances shows a fluctuating but overall increasing trend. It is worth noting that PaddlePaddle exhibits a significant peak at the 8th version, but experiences a sharp decrease with number of instances of \textit{UNE} (108 to 57), \textit{LLF} (110 to 61), and \textit{LILBC} (78 to 41) from the 8th to the 9th version. 

On average, \textit{LLF} (except for MindSpore) and \textit{UNE} (except for Chainer) show more noticeable variations. Additionally, the numbers of \textit{LSD} and \textit{LILBC} instances also show significant growth in some DLFs. For instance, in PyTorch, the number of \textit{LSD} instances has grown the fastest, more than doubling; in MindSpore, the number of \textit{LILBC} instances has grown the fastest, more than three times the original number.

\begin{framed}
\noindent\textbf{Answer to RQ3: }The number of ILDS instances in most DLFs shows an overall upward trend, and \textit{LLF} and \textit{UNE} exhibit more dynamical evolution overall.
\end{framed}

\subsection{Discussion}\label{s:6}
\subsubsection{Understanding of Study Results}
\textbf{On RQ1 results.}
Since our detection method relies on ILCMs, and different projects use different ILCMs, this will lead to variations in detection results. Besides, the difference in the distribution of the 7 ILDS in the 5 DLFs is mainly affected by two aspects. 
\textbf{(1) Developer coding practice preference.} We observed that developers working on MindSpore tend to use classes instead of functions when writing C++ extension modules. This preference may lead to the introduction of more \textit{LILBC} instances.
In PaddlePaddle, we observed that the developer bound all native entities into an extension module, leading to frequent calls to this module during inter-language interactions. As a result, some unused native entities may be overlooked, similar to a frequently used class where certain functions are never utilized.
\textbf{(2) Developer programming standards.}
Furthermore, we found that although developers' programming habits are a decisive factor in the occurrence of ILDS, in some DLFs, developers may pay more attention to programming standards. For example, in TensorFlow and PyTorch, developers tend to use lambda functions when binding C++ functions in extension modules, which leads to a large number of \textit{UNE} instances. As shown in Table~\ref{tab:t5}, \textit{UNE} instances in TensorFlow account for 67.71\%, more than double that of PyTorch (30.90\%). However, we found that in PyTorch, 78.71\% of the bindings use lambda functions when binding C++ functions, higher than TensorFlow's 70.72\%. This indicates that the management of lambda function usage is more standardized in PyTorch.
\textbf{(3) The structures of the DLFs.} 
We found that in PyTorch, pybind11 is used mainly by Caffe, which is another DLF that was merged into PyTorch in 2018. The 119 native functions in Caffe are bound in 4 modules, and these functions use Lambda expressions extensively, resulting in a high density of \textit{LLF} in PyTorch. Furthermore, PyTorch relies heavily on the Python/C API, and our analysis identified 393 functions directly bound through the Python/C API, which is about 28 times that of TensorFlow. This also contributes to a higher occurrence of \textit{LSD} in PyTorch.

According to Table~\ref{tab:t5}, some of the smells have relatively few instances. We found that the low number of instances of LREC and NURP is related to the limited number of scenarios that meet the requirements for these smells. For example, LREC indicates the lack of rigorous error checking during module initialization, which only occurs in inter-language communication using the Python/C API. Based on the results in Table~\ref{tab:t1}, only TensorFlow, PyTorch, and PaddlePaddle meet this condition. Moreover, as we mentioned in the previous paragraph, inter-language bindings in DLFs are relatively concentrated. For instance, in PaddlePaddle, there is only one main extension module, which leads to fewer occurrences of such smells. The number of EILC instances depends on the threshold definition. Reducing this threshold will result in a large increase in EILC instances. The exact threshold should be set according to practical needs. A higher threshold indicates a greater degree of coupling between components. Although these smells appear less frequently in DLFs, this does not mean they are unimportant. Similar to EILC, Groot et al. further studied eight types of dependencies between components in multi-language systems and elaborated on their potential harms~\citep{groot2024catalog}.

In addition, we found that ILDS appears more frequently on the C/C++ side, primarily for two reasons. First, in multi-language programming for DLFs, C/C++ typically provides function interfaces, while Python is mainly responsible for calling these interfaces. Second, in both the Python/C API and pybind11, the inter-language binding process occurs on the C/C++ side, making it more likely for ILDS to occur in C/C++.

\textbf{On RQ2 results.} 
We found that fixes for ILDS are often accompanied by updates to the DLF structure. For example, PaddlePaddle exhibited a significantly higher fix ratio of 22.73\% for \textit{LLF} than 3.11\% of TensorFlow and 1.37\% of PyTorch. This discrepancy can be attributed to that our dataset included PaddlePaddle's first major version (1.8.5) and its last major version (2.0.0) before the study. By analyzing the detection results and release notes of PaddlePaddle, we found that after this major version update (from 1.8.5 to 2.0.0), PaddlePaddle's internal architecture underwent substantial changes. Specifically, many API names were altered, especially with improvements in PaddlePaddle's dynamic graph capabilities, resulting in significant adjustments to related dynamic graph processing functions, and consequently affecting the proportion of ILDS fixes. Furthermore, \textit{LREC}, and \textit{NURP} were not fixed at all, which suggests that these ILDS have persisted in the DLF despite of their initially low occurrences. We hope that our report on such ILDS can attract the developers' attention from the DLFs.

\textbf{On RQ3 results.}
As the oldest DLF, Chainer has seen minimal updates over the past three years, therefore the number of ILDS instances in Chainer has remained relatively stable. In contrast, TensorFlow, PyTorch, PaddlePaddle, and MindSpore all exhibit an upward trend in the number of ILDS instances, which implies that even though some ILDS instances were fixed, more new ILDS instances were still introduced in later versions. The possible reason is that as the version evolves, the system's functionalities are continuously improved, making the codebase more complex, and thus increasing the probability of ILDS introduction. Meanwhile, ILDS as a type of technical debt can be incurred by inappropriate technical decisions~\citep{li2015systematic}. 

In particular, the number of ILDS instances of certain DLFs peaks during the version evolution process. For example, the number of ILDS instances reached the highest point at the 8th version of PaddlePaddle, and  dropped significantly in the subsequent versions. Our analysis of the ILDS data detected by \textsc{CPsmell} indicates that the 9th version of PaddlePaddle reduced the use of functions, classes, and modules for interaction by approximately a half compared with the 8th version. Combined with the updated data from the PaddlePaddle version~\citep{paddlepaddle_discuss}, it becomes apparent that the transition from the 8th to the 9th version involved relatively fewer additions of new inter-language APIs and instead focused predominantly on updates and issue resolutions. This means that these ILDS instances affect the update and maintenance of the system. The overall upward trend of ILDS instances also shows that the maintainability of the DLFs still needs to be improved.

We also observed that updates to DLFs often involve the addition of inter-language APIs, supported by native C/C++ functions that are usually bound to extension modules in the form of Lambda expressions. This leads to an increase of \textit{LLF} instances. At the same time, updates to DLFs can also result in the deprecation of old APIs, leading to the occurrence of \textit{UNE} instances (as depicted in Figure \ref{fig:f5}, particularly evident in PaddlePaddle). This can also partially explain why the \textit{LLF} and \textit{UNE} instances account for a relatively large proportion in RQ1.

\vspace*{-3pt}
\section{Limitations}\label{sec_limitation}

\textbf{Limitations of the ILDS extraction process.}
One limitation is related to the data sources of the ILDS extraction process (see Figure~\ref{fig:f1}). We collected data from technical documents, GitHub issues, relevant literature, and technical community, using keyword-based mining. This approach could potentially introduce biases into our results. To mitigate potential biases, we specifically focused on official technical documentation and issues effectively resolved within active communities, and we took validation examples for the defined ILDS from the source code of 5 popular DLFs,  which are somewhat representative of the deep learning field. 

\textbf{Limitations of the implementation and validation of ILDS detection.}
(1) Since the threshold values in our defined ILDS identification rules are based on experience and relevant literature~\citep{abidi2021multi, chen2016detecting,lippert2006refactoring}, this may not be suitable for all multi-language software systems written in Python and C/C++. Thus, we allow flexible adjustment in the configuration file of \textsc{CPsmell} to improve its generailizability. 
(2) \textsc{CPsmell} only supports three relatively mature ILCMs for Python and C/C++, i.e., Python/C API, pybind11, and ctypes. In addition, these ILDS are proposed in this work and there is no other detection tool that can be compared with \textsc{CPsmell}.
(3) Another limitation is related to the manual validation of the \textsc{CPsmell} tool. Due to the imbalance of the number of the identified instances for different ILDS, we could not validate many instances of \textit{LREC}, \textit{NURP}, and \textit{EILC}, as shown in Table \ref{tab:t10}. 

\textbf{Limitations of the empirical study.}
(1) One limitation comes from the  \textsc{CPsmell} tool, which detection accuracy is 98.17\%, meaning that there are still 1.83\% ILDS instances being incorrectly identified. Given the high accuracy of \textsc{CPsmell}, the negative impact on the empirical study results is rather limited. 
(2) Our empirical study only validated the proposed ILDS in multi-language DLFs, and a more comprehensive validation of the proposed ILDS is needed for a larger range of multi-language software systems (in addition to DLFs) written in Python and C/C++.


\vspace*{-3pt}
\section{Implications}\label{sec_implication}
\subsection{Implications for Practitioners}
\textbf{The proposed ILDS are helpful in improving the code quality in multi-language software systems.} For example, we recommend developers to define long Lambda functions separately when using pybind11 to write C++ extension modules. This practice improves the clarity of the entire binding process. In addition, when using the Python/C API, it is essential to set the bound functions as static to mitigate namespace pollution. Readers can find more actionable suggestions on code quality improvement in Section \ref{sec_ILDS}, where we provide a refactoring recommendation for each ILDS.

\textbf{DLF developers are recommended to pay more attention to the reported ILDS with higher possibility to occur in DLFs.} We investigated the distribution and evolution process of different ILDS in DLFs. The results are beneficial to the developers' understanding on which ILDS are more likely to occur. For example, the proportion of the \textit{LSD} instances in PyTorch has reached 41.2\%, and furthermore the number of \textit{LSD} instances has more than doubled as shown in Figure~\ref{fig:f5}. Therefore, we recommend maintainers of PyTorch to pay more attention to \textit{LSD}. Additionally, our research found that \textit{LLF} and \textit{UNE} are more likely to occur and have higher proportions (both exceed 25\%) compared to other ILDS. This is closely related to the increase of inter-language APIs during version evolution. Thus, we recommend developers to pay extra attention to \textit{LLF} and \textit{UNE} when adding new inter-language APIs.

\textbf{DLF developers are advised to invest effort in fixing historical ILDS instances.} In RQ2 and RQ3, we found that some ILDS instances were fixed during the evolution of versions, which indicates that the ILDS instances indeed had negative impacts on the system maintainability. 
Although \textit{LREC}, and \textit{NURP} were not fixed during the 3-year evolution, as shown in Table~\ref{tab:t9}, it does not mean that these ILDS do not need to be fixed. 
For instance, \textit{LREC} instances were not effectively fixed in the 3 years, but at least one instance had been resolved before as described in the Section~\ref{sec_LREC}, where the initialization function of an extension module in PyTorch was refactored. 

\subsection{Implications for Researchers}
\textbf{There is a need for research on ILDS for an extensive PL combinations.} Most of the related work has focused on maintainability issues of the multi-language software systems in Java and C/C++~\citep{abidi2019anti, abidi2019code, abidi2021multi}. In contrast, our work investigates ILDS in 5 DLFs, a typical set of multi-language software systems written in Python and C/C++, providing researchers with a more comprehensive understanding on ILDS in a certain application domain.
Further research on ILDS for an extensive PL combinations will deepen and broaden the understanding on ILDS in different domains.

\textbf{Research on multi-language software can benefit from the perspective of using ILCMs as bridges to analyze multi-language software in our study.} We detected and analyzed the ILDS in DLFs based on three widely-used ILCMs, namely Python/C API, pybind11, and ctypes, which provides a practical perspective to analyze the ILDS that may arise from the selection of different ILCMs for any PL combinations.

\textbf{\textsc{CPsmell} can facilitate the research on inter-language analysis on software written in the combination of Python and C/C++.} Nowadays, besides DLFs, the combination of Python and C/C++ is gaining increasing popularity, which even surpasses the combination of Java and C/C++~\citep{li2024multilingual}. However, there is a very limited number of inter-language analysis methods for Python and C/C++. Thus, we developed the \textsc{CPsmell} tool specifically for this PL combination, providing convenience for other researchers.


\section{Conclusions}\label{sec_conclusion}
In this work, we proposed 7 inter-language design smells (ILDS) for multi-language deep learning frameworks (DLFs) written in Python and C/C++, and extracted the detection rule for each ILDS based on three widely used inter-language communication mechanisms between Python and C/C++ in such DLFs. We then implemented the ILDS detection in the \textsc{CPsmell} tool, with an accuracy of 98.17\% in our manual validation in the DLFs. Our empirical study on multi-language DLFs reveals significant differences in the distribution of ILDS, with ILDS \textit{Long Lambda Function For Inter-language Binding} and \textit{Unused Native Entity} most widely distributed. We also found that some ILDS are more likely to be fixed during version evolution, e.g., \textit{Excessive Inter-Language Communication}, \textit{Large Inter-language Binding Class}, and \textit{Unused Native Entity}. In addition, by analyzing 50 versions of the DLFs, we found that the number of most ILDS instances shows an overall upward trend, which makes the DLFs more difficult to maintain.

Our proposed ILDS are defined and identified based on inter-language communication mechanisms between Python and C/C++, and are thus not specifically limited to DLFs. In the next step, we will validate the ILDS to multi-language systems involving Python and C/C++ other than DLFs to enhance the generalizability of the proposed ILDS. We also plan to explore more inter-language communication methods, such as message passing and glue code, and investigate combinations of other programming languages to identify more ILDS.

\section*{Data availability}
We have shared the link to our dataset in the reference~\citep{cpsmell}.

\section*{Acknowledgments}
This work was funded by the National Natural Science Foundation of China under Grant Nos. 62176099, 62172311, and 62402499, and the Knowledge Innovation Program of Wuhan-Shuguang Project under Grant No. 2022010801020280.

\printcredits
\bibliographystyle{cas-model2-names}

\bibliography{reference}

\begin{thebibliography}{60}
\expandafter\ifx\csname natexlab\endcsname\relax\def\natexlab#1{#1}\fi
\providecommand{\url}[1]{\texttt{#1}}
\providecommand{\href}[2]{#2}
\providecommand{\path}[1]{#1}
\providecommand{\DOIprefix}{doi:}
\providecommand{\ArXivprefix}{arXiv:}
\providecommand{\URLprefix}{URL: }
\providecommand{\Pubmedprefix}{pmid:}
\providecommand{\doi}[1]{\href{http://dx.doi.org/#1}{\path{#1}}}
\providecommand{\Pubmed}[1]{\href{pmid:#1}{\path{#1}}}
\providecommand{\bibinfo}[2]{#2}
\ifx\xfnm\relax \def\xfnm[#1]{\unskip,\space#1}\fi
\bibitem[{Abadi et~al.(2016)Abadi, Barham, Chen, Chen, Davis, Dean, Devin, Ghemawat, Irving, Isard et~al.}]{abadi2016tensorflow}
\bibinfo{author}{Abadi, M.}, \bibinfo{author}{Barham, P.}, \bibinfo{author}{Chen, J.}, \bibinfo{author}{Chen, Z.}, \bibinfo{author}{Davis, A.}, \bibinfo{author}{Dean, J.}, \bibinfo{author}{Devin, M.}, \bibinfo{author}{Ghemawat, S.}, \bibinfo{author}{Irving, G.}, \bibinfo{author}{Isard, M.}, et~al., \bibinfo{year}{2016}.
\newblock \bibinfo{title}{{TensorFlow}: a system for large-scale machine learning}, in: \bibinfo{booktitle}{Proceedings of the 12th USENIX symposium on operating systems design and implementation (OSDI)}, \bibinfo{address}{Savannah, GA, USA}. pp. \bibinfo{pages}{265--283}.
\bibitem[{Abidi et~al.(2019a)Abidi, Grichi, Khomh and Gu{\'e}h{\'e}neuc}]{abidi2019code}
\bibinfo{author}{Abidi, M.}, \bibinfo{author}{Grichi, M.}, \bibinfo{author}{Khomh, F.}, \bibinfo{author}{Gu{\'e}h{\'e}neuc, Y.G.}, \bibinfo{year}{2019}a.
\newblock \bibinfo{title}{Code smells for multi-language systems}, in: \bibinfo{booktitle}{Proceedings of the 24th European conference on pattern languages of programs (EuroPLoP)}, \bibinfo{publisher}{ACM}, \bibinfo{address}{Irsee, Germany}. pp. \bibinfo{pages}{1--13}.
\bibitem[{Abidi et~al.(2019b)Abidi, Khomh and Gu{\'e}h{\'e}neuc}]{abidi2019anti}
\bibinfo{author}{Abidi, M.}, \bibinfo{author}{Khomh, F.}, \bibinfo{author}{Gu{\'e}h{\'e}neuc, Y.G.}, \bibinfo{year}{2019}b.
\newblock \bibinfo{title}{Anti-patterns for multi-language systems}, in: \bibinfo{booktitle}{Proceedings of the 24th European conference on pattern languages of programs (EuroPLoP)}, \bibinfo{publisher}{ACM}, \bibinfo{address}{Irsee, Germany}. pp. \bibinfo{pages}{1--14}.
\bibitem[{Abidi et~al.(2021)Abidi, Rahman, Openja and Khomh}]{abidi2021multi}
\bibinfo{author}{Abidi, M.}, \bibinfo{author}{Rahman, M.S.}, \bibinfo{author}{Openja, M.}, \bibinfo{author}{Khomh, F.}, \bibinfo{year}{2021}.
\newblock \bibinfo{title}{Are multi-language design smells fault-prone? an empirical study}.
\newblock \bibinfo{journal}{ACM Transactions on Software Engineering and Methodology} \bibinfo{volume}{30}, \bibinfo{pages}{1--56}.
\bibitem[{Baidu(2023a)}]{paddlepaddle_eilc}
\bibinfo{author}{Baidu}, \bibinfo{year}{2023}a.
\newblock \bibinfo{title}{Paddlepaddle}.
\newblock \bibinfo{note}{\href{https://github.com/PaddlePaddle/Paddle/blob/v2.4.2/python/paddle/fluid/dygraph/varbase_patch_methods.py}{https://github.com/PaddlePaddle/Paddle/blob/v2.4.2/python/paddle/fluid/dygraph/varbase\_patch\_methods.py}}.
\bibitem[{Baidu(2023b)}]{paddlepaddle_discuss}
\bibinfo{author}{Baidu}, \bibinfo{year}{2023}b.
\newblock \bibinfo{title}{Paddlepaddle releases}.
\newblock \bibinfo{note}{\href{https://github.com/PaddlePaddle/Paddle/releases}{https://github.com/PaddlePaddle/Paddle/releases}}.
\bibitem[{Chen et~al.(2016)Chen, Chen, Ma and Xu}]{chen2016detecting}
\bibinfo{author}{Chen, Z.}, \bibinfo{author}{Chen, L.}, \bibinfo{author}{Ma, W.}, \bibinfo{author}{Xu, B.}, \bibinfo{year}{2016}.
\newblock \bibinfo{title}{Detecting code smells in python programs}, in: \bibinfo{booktitle}{Proceedings of the 2016 international conference on Software Analysis, Testing and Evolution (SATE)}, \bibinfo{publisher}{IEEE}, \bibinfo{address}{Kunming, China}. pp. \bibinfo{pages}{18--23}.
\bibitem[{Facebook(2018)}]{pytorch_lrec0.3.1}
\bibinfo{author}{Facebook}, \bibinfo{year}{2018}.
\newblock \bibinfo{title}{Pytorch}.
\newblock \bibinfo{note}{\href{https://github.com/pytorch/pytorch/blob/v0.3.1/torch/csrc/Size.cpp}{https://github.com/pytorch/pytorch/blob/v0.3.1/torch/csrc/Size.cpp}}.
\bibitem[{Facebook(2019)}]{pytorch_lrec1.1.0}
\bibinfo{author}{Facebook}, \bibinfo{year}{2019}.
\newblock \bibinfo{title}{Pytorch}.
\newblock \bibinfo{note}{\href{https://github.com/pytorch/pytorch/blob/v1.1.0/torch/csrc/Size.cpp}{https://github.com/pytorch/pytorch/blob/v1.1.0/torch/csrc/Size.cpp}}.
\bibitem[{Facebook(2022)}]{pytorch_uem}
\bibinfo{author}{Facebook}, \bibinfo{year}{2022}.
\newblock \bibinfo{title}{Pytorch}.
\newblock \bibinfo{note}{\href{https://github.com/pytorch/pytorch/blob/v1.11.0/torch/csrc/deploy/test\_deploy\_lib.cpp}{https://github.com/pytorch/pytorch/blob/v1.11.0/torch/csrc/deploy/test\_deploy\_lib.cpp}}.
\bibitem[{Facebook(2023a)}]{pytorch_lsd}
\bibinfo{author}{Facebook}, \bibinfo{year}{2023}a.
\newblock \bibinfo{title}{Pytorch}.
\newblock \bibinfo{note}{\href{https://github.com/pytorch/pytorch/blob/v2.0.0/torch/csrc/Module.cpp}{https://github.com/pytorch/pytorch/blob/v2.0.0/torch/csrc/Module.cpp}}.
\bibitem[{Facebook(2023b)}]{pytorch_nurp}
\bibinfo{author}{Facebook}, \bibinfo{year}{2023}b.
\newblock \bibinfo{title}{Pytorch}.
\newblock \bibinfo{note}{\href{https://github.com/pytorch/pytorch/blob/v2.0.1/torch/cuda/__init__.py}{https://github.com/pytorch/pytorch/blob/v2.0.1/torch/cuda/\_\_init\_\_.py}}.
\bibitem[{Facebook(2023c)}]{pytorch_discuss1}
\bibinfo{author}{Facebook}, \bibinfo{year}{2023}c.
\newblock \bibinfo{title}{Pytorch}.
\newblock \bibinfo{note}{\href{https://github.com/pytorch/pytorch/blob/v2.0.1/torch/_inductor/codecache.py}{https://github.com/pytorch/pytorch/blob/v2.0.1/torch/\_inductor/codecache.py}}.
\bibitem[{Facebook(2023d)}]{pytorch_discuss2}
\bibinfo{author}{Facebook}, \bibinfo{year}{2023}d.
\newblock \bibinfo{title}{Pytorch}.
\newblock \bibinfo{note}{\href{https://github.com/pytorch/pytorch/blob/v2.0.1/torch/utils/benchmark/utils/valgrind_wrapper/timer_interface.py}{https://github.com/pytorch/pytorch/blob/v2.0.1/torch/utils/benchmark/utils/valgrind\_wrapper/timer\_interface.py}}.
\bibitem[{Foundation(2019a)}]{ctypes}
\bibinfo{author}{Foundation, P.S.}, \bibinfo{year}{2019}a.
\newblock \bibinfo{title}{ctypes — a foreign function library for python}.
\newblock \bibinfo{note}{\href{https://docs.python.org/3/library/ctypes.html}{https://docs.python.org/3/library/ctypes.html}}.
\bibitem[{Foundation(2019b)}]{python/c_api}
\bibinfo{author}{Foundation, P.S.}, \bibinfo{year}{2019}b.
\newblock \bibinfo{title}{Extending and embedding the python interpreter}.
\newblock \bibinfo{note}{\href{https://docs.python.org/zh-cn/3.8/extending/index.html\#extending-index}{https://docs.python.org/zh-cn/3.8/extending/index.html\#extending-index}}.
\bibitem[{Fowler(2018)}]{fowler2018refactoring}
\bibinfo{author}{Fowler, M.}, \bibinfo{year}{2018}.
\newblock \bibinfo{title}{Refactoring}.
\newblock \bibinfo{edition}{2nd} ed., \bibinfo{publisher}{Addison-Wesley Professional}.
\bibitem[{Gesi et~al.(2022)Gesi, Liu, Li, Ahmed, Nagappan, Lo, de~Almeida, Kochhar and Bao}]{gesi2022code}
\bibinfo{author}{Gesi, J.}, \bibinfo{author}{Liu, S.}, \bibinfo{author}{Li, J.}, \bibinfo{author}{Ahmed, I.}, \bibinfo{author}{Nagappan, N.}, \bibinfo{author}{Lo, D.}, \bibinfo{author}{de~Almeida, E.S.}, \bibinfo{author}{Kochhar, P.S.}, \bibinfo{author}{Bao, L.}, \bibinfo{year}{2022}.
\newblock \bibinfo{title}{Code smells in machine learning systems}.
\newblock \bibinfo{journal}{arXiv preprint arXiv:2203.00803} .
\bibitem[{Google(2023a)}]{google_style}
\bibinfo{author}{Google}, \bibinfo{year}{2023}a.
\newblock \bibinfo{title}{Google python style guide}.
\newblock \bibinfo{note}{\href{https://google.github.io/styleguide/pyguide.html}{https://google.github.io/styleguide/pyguide.html}}.
\bibitem[{Google(2023b)}]{tensorflow_llf}
\bibinfo{author}{Google}, \bibinfo{year}{2023}b.
\newblock \bibinfo{title}{Tensorflow}.
\newblock \bibinfo{note}{\href{https://github.com/tensorflow/tensorflow/blob/v2.12.0/tensorflow/python/tfe\_wrapper.cc}{https://github.com/tensorflow/tensorflow/blob/v2.12.0/tensorflow/python/tfe\_wrapper.cc}}.
\bibitem[{Google(2023c)}]{tensorflow_lilbc}
\bibinfo{author}{Google}, \bibinfo{year}{2023}c.
\newblock \bibinfo{title}{Tensorflow}.
\newblock \bibinfo{note}{\href{https://github.com/tensorflow/tensorflow/blob/v2.12.0/tensorflow/compiler/xla/python/xla.cc}{https://github.com/tensorflow/tensorflow/blob/v2.12.0/tensorflow/compiler/xla/python/xla.cc}}.
\bibitem[{Grichi(2020)}]{grichi2020towards}
\bibinfo{author}{Grichi, M.}, \bibinfo{year}{2020}.
\newblock \bibinfo{title}{Towards Understanding Modern Multi-Language Software\\ Systems}.
\newblock Ph.D. thesis. Ecole Polytechnique, Montreal, Canada.
\bibitem[{Grichi et~al.(2021)Grichi, Abidi, Jaafar, Eghan and Adams}]{grichi2021impact}
\bibinfo{author}{Grichi, M.}, \bibinfo{author}{Abidi, M.}, \bibinfo{author}{Jaafar, F.}, \bibinfo{author}{Eghan, E.E.}, \bibinfo{author}{Adams, B.}, \bibinfo{year}{2021}.
\newblock \bibinfo{title}{On the impact of interlanguage dependencies in multilanguage systems empirical case study on java native interface applications (jni)}.
\newblock \bibinfo{journal}{IEEE Transactions on Reliability} \bibinfo{volume}{70}, \bibinfo{pages}{428--440}.
\bibitem[{Grichi et~al.(2020)Grichi, Eghan and Adams}]{grichi2020MLF}
\bibinfo{author}{Grichi, M.}, \bibinfo{author}{Eghan, E.E.}, \bibinfo{author}{Adams, B.}, \bibinfo{year}{2020}.
\newblock \bibinfo{title}{On the impact of multi-language development in machine learning frameworks}, in: \bibinfo{booktitle}{Proceedings of the 36th IEEE International Conference on Software Maintenance and Evolution (ICSME)}, \bibinfo{organization}{IEEE}. pp. \bibinfo{pages}{546--556}.
\bibitem[{Groot et~al.(2024)Groot, Venegas, Laz{\u{a}}r and Kr{\"u}ger}]{groot2024catalog}
\bibinfo{author}{Groot, T.}, \bibinfo{author}{Venegas, L.O.}, \bibinfo{author}{Laz{\u{a}}r, B.}, \bibinfo{author}{Kr{\"u}ger, J.}, \bibinfo{year}{2024}.
\newblock \bibinfo{title}{A catalog of unintended software dependencies in multi-lingual systems at asml}, in: \bibinfo{booktitle}{Proceedings of the 46th International Conference on Software Engineering: Software Engineering in Practice (ICSE-SEIP)}, \bibinfo{organization}{ACM}, \bibinfo{address}{New York, NY, USA}. pp. \bibinfo{pages}{240--–251}.
\bibitem[{Hayworth and Maletic(2022)}]{pylibsrcml}
\bibinfo{author}{Hayworth, M.}, \bibinfo{author}{Maletic, J.I.}, \bibinfo{year}{2022}.
\newblock \bibinfo{title}{pylibsrcml}.
\newblock \bibinfo{note}{\href{https://github.com/srcML/pylibsrcml}{https://github.com/srcML/pylibsrcml}}.
\bibitem[{Hu and Zhang(2020)}]{hu2020python}
\bibinfo{author}{Hu, M.}, \bibinfo{author}{Zhang, Y.}, \bibinfo{year}{2020}.
\newblock \bibinfo{title}{The python/c api: evolution, usage statistics, and bug patterns}, in: \bibinfo{booktitle}{Proceedings of the 27th International Conference on Software Analysis, Evolution and Reengineering (SANER)}, \bibinfo{organization}{IEEE}. pp. \bibinfo{pages}{532--536}.
\bibitem[{Huawei(2023)}]{mindspore_lilbc}
\bibinfo{author}{Huawei}, \bibinfo{year}{2023}.
\newblock \bibinfo{title}{Mindspore}.
\newblock \bibinfo{note}{\href{https://github.com/mindspore-ai/mindspore/blob/v2.0.0/mindspore/ccsrc/pybind_api/ir/tensor_py.cc}{https://github.com/mindspore-ai/mindspore/blob/v2.0.0/mindspore/ccsrc/pybind\_api/ir/tensor\_py.cc}}.
\bibitem[{Huawei Technologies~Co.(2022)}]{huawei2022huawei}
\bibinfo{author}{Huawei Technologies~Co., L.}, \bibinfo{year}{2022}.
\newblock \bibinfo{title}{Huawei mindspore ai development framework}, in: \bibinfo{booktitle}{Artificial Intelligence Technology}. \bibinfo{publisher}{Springer}, pp. \bibinfo{pages}{137--162}.
\bibitem[{Jakob(2017)}]{pybind11}
\bibinfo{author}{Jakob, W.}, \bibinfo{year}{2017}.
\newblock \bibinfo{title}{pybind11 — seamless operability between c++11 and python}.
\newblock \bibinfo{note}{\href{https://pybind11.readthedocs.io/en/stable/index.html}{https://pybind11.readthedocs.io/en/stable/index.html}}.
\bibitem[{Jebnoun et~al.(2020)Jebnoun, Ben~Braiek, Rahman and Khomh}]{jebnoun2020scent}
\bibinfo{author}{Jebnoun, H.}, \bibinfo{author}{Ben~Braiek, H.}, \bibinfo{author}{Rahman, M.M.}, \bibinfo{author}{Khomh, F.}, \bibinfo{year}{2020}.
\newblock \bibinfo{title}{The scent of deep learning code: An empirical study}, in: \bibinfo{booktitle}{Proceedings of the 17th International Conference on Mining Software Repositories (MSR)}, pp. \bibinfo{pages}{420--430}.
\bibitem[{Kochhar et~al.(2016)Kochhar, Wijedasa and Lo}]{kochhar2016large}
\bibinfo{author}{Kochhar, P.S.}, \bibinfo{author}{Wijedasa, D.}, \bibinfo{author}{Lo, D.}, \bibinfo{year}{2016}.
\newblock \bibinfo{title}{A large scale study of multiple programming languages and code quality}, in: \bibinfo{booktitle}{Proceedings of the 23rd IEEE International Conference on Software Analysis, Evolution, and Reengineering (SANER)}, \bibinfo{organization}{IEEE}. pp. \bibinfo{pages}{563--573}.
\bibitem[{Kullbach et~al.(1998)Kullbach, Winter, Dahm and Ebert}]{kullbach1998program}
\bibinfo{author}{Kullbach, B.}, \bibinfo{author}{Winter, A.}, \bibinfo{author}{Dahm, P.}, \bibinfo{author}{Ebert, J.}, \bibinfo{year}{1998}.
\newblock \bibinfo{title}{Program comprehension in multi-language systems}, in: \bibinfo{booktitle}{Proceedings of the 5th Working Conference on Reverse Engineering (WCRE)}, \bibinfo{organization}{IEEE}. pp. \bibinfo{pages}{135--143}.
\bibitem[{Li et~al.(2022a)Li, Li and Cai}]{li2022vulnerability}
\bibinfo{author}{Li, W.}, \bibinfo{author}{Li, L.}, \bibinfo{author}{Cai, H.}, \bibinfo{year}{2022}a.
\newblock \bibinfo{title}{On the vulnerability proneness of multilingual code}, in: \bibinfo{booktitle}{Proceedings of the 30th ACM Joint European Software Engineering Conference and Symposium on the Foundations of Software Engineering (ESEC/FSE)}, \bibinfo{organization}{ACM}. pp. \bibinfo{pages}{847--859}.
\bibitem[{Li et~al.(2022b)Li, Li and Cai}]{li2022polyfax}
\bibinfo{author}{Li, W.}, \bibinfo{author}{Li, L.}, \bibinfo{author}{Cai, H.}, \bibinfo{year}{2022}b.
\newblock \bibinfo{title}{Polyfax: a toolkit for characterizing multi-language software}, in: \bibinfo{booktitle}{Proceedings of the 30th ACM Joint European Software Engineering Conference and Symposium on the Foundations of Software Engineering (ESEC/FSE)}, \bibinfo{organization}{ACM}. pp. \bibinfo{pages}{1662--1666}.
\bibitem[{Li et~al.(2024a)Li, Marino, Yang, Meng, Li and Cai}]{li2024multilingual}
\bibinfo{author}{Li, W.}, \bibinfo{author}{Marino, A.}, \bibinfo{author}{Yang, H.}, \bibinfo{author}{Meng, N.}, \bibinfo{author}{Li, L.}, \bibinfo{author}{Cai, H.}, \bibinfo{year}{2024}a.
\newblock \bibinfo{title}{How are multilingual systems constructed: Characterizing language use and selection in open-source multilingual software}.
\newblock \bibinfo{journal}{ACM Transactions on Software Engineering and Methodology} \bibinfo{volume}{33}, \bibinfo{pages}{1--46}.
\bibitem[{Li et~al.(2022c)Li, Ming, Luo and Cai}]{li2022polycruise}
\bibinfo{author}{Li, W.}, \bibinfo{author}{Ming, J.}, \bibinfo{author}{Luo, X.}, \bibinfo{author}{Cai, H.}, \bibinfo{year}{2022}c.
\newblock \bibinfo{title}{{PolyCruise}: A cross-language dynamic information flow analysis}, in: \bibinfo{booktitle}{Proceedings of the 31st USENIX Security Symposium (USENIX Security)}, pp. \bibinfo{pages}{2513--2530}.
\bibitem[{Li et~al.(2015)Li, Avgeriou and Liang}]{li2015systematic}
\bibinfo{author}{Li, Z.}, \bibinfo{author}{Avgeriou, P.}, \bibinfo{author}{Liang, P.}, \bibinfo{year}{2015}.
\newblock \bibinfo{title}{A systematic mapping study on technical debt and its management}.
\newblock \bibinfo{journal}{Journal of Systems and Software} \bibinfo{volume}{101}, \bibinfo{pages}{193--220}.
\bibitem[{Li et~al.(2023a)Li, Wang, Wang, Liang, Mo and Li}]{li2023DLF}
\bibinfo{author}{Li, Z.}, \bibinfo{author}{Wang, S.}, \bibinfo{author}{Wang, W.}, \bibinfo{author}{Liang, P.}, \bibinfo{author}{Mo, R.}, \bibinfo{author}{Li, B.}, \bibinfo{year}{2023}a.
\newblock \bibinfo{title}{Understanding bugs in multi-language deep learning frameworks}, in: \bibinfo{booktitle}{Proceedings of the 31st International Conference on Program Comprehension (ICPC)}, \bibinfo{organization}{IEEE}. pp. \bibinfo{pages}{328--338}.
\bibitem[{Li et~al.(2023b)Li, Wang, Wang, Liang and Mo}]{li2023understanding}
\bibinfo{author}{Li, Z.}, \bibinfo{author}{Wang, W.}, \bibinfo{author}{Wang, S.}, \bibinfo{author}{Liang, P.}, \bibinfo{author}{Mo, R.}, \bibinfo{year}{2023}b.
\newblock \bibinfo{title}{Understanding resolution of multi-language bugs: An empirical study on apache projects}, in: \bibinfo{booktitle}{Proceedings of the 17th ACM/IEEE International Symposium on Empirical Software Engineering and Measurement (ESEM)}, \bibinfo{organization}{IEEE}. pp. \bibinfo{pages}{1--11}.
\bibitem[{Li et~al.(2024b)Li, Zhang, Wang, Liang, Mo, Tan and Liu}]{cpsmell}
\bibinfo{author}{Li, Z.}, \bibinfo{author}{Zhang, X.}, \bibinfo{author}{Wang, W.}, \bibinfo{author}{Liang, P.}, \bibinfo{author}{Mo, R.}, \bibinfo{author}{Tan, J.}, \bibinfo{author}{Liu, H.}, \bibinfo{year}{2024}b.
\newblock \bibinfo{title}{Cpsmell}.
\newblock \bibinfo{note}{\href{https://github.com/ILDSOFDLF/cpsmell}{https://github.com/ILDSOFDLF/cpsmell}}.
\bibitem[{Lippert and Roock(2006)}]{lippert2006refactoring}
\bibinfo{author}{Lippert, M.}, \bibinfo{author}{Roock, S.}, \bibinfo{year}{2006}.
\newblock \bibinfo{title}{Refactoring in large software projects: performing complex restructurings successfully}.
\newblock \bibinfo{publisher}{John Wiley \& Sons}.
\bibitem[{Ma et~al.(2019)Ma, Yu, Wu and Wang}]{ma2019paddlepaddle}
\bibinfo{author}{Ma, Y.}, \bibinfo{author}{Yu, D.}, \bibinfo{author}{Wu, T.}, \bibinfo{author}{Wang, H.}, \bibinfo{year}{2019}.
\newblock \bibinfo{title}{Paddlepaddle: An open-source deep learning platform from industrial practice}.
\newblock \bibinfo{journal}{Frontiers of Data and Domputing} \bibinfo{volume}{1}, \bibinfo{pages}{105--115}.
\bibitem[{Mayer et~al.(2017)Mayer, Kirsch and Le}]{mayer2017multi}
\bibinfo{author}{Mayer, P.}, \bibinfo{author}{Kirsch, M.}, \bibinfo{author}{Le, M.A.}, \bibinfo{year}{2017}.
\newblock \bibinfo{title}{On multi-language software development, cross-language links and accompanying tools: a survey of professional software developers}.
\newblock \bibinfo{journal}{Journal of Software Engineering Research and Development} \bibinfo{volume}{5}, \bibinfo{pages}{1--33}.
\bibitem[{Moha et~al.(2009)Moha, Gu{\'e}h{\'e}neuc, Duchien and Le~Meur}]{moha2009decor}
\bibinfo{author}{Moha, N.}, \bibinfo{author}{Gu{\'e}h{\'e}neuc, Y.G.}, \bibinfo{author}{Duchien, L.}, \bibinfo{author}{Le~Meur, A.F.}, \bibinfo{year}{2009}.
\newblock \bibinfo{title}{Decor: A method for the specification and detection of code and design smells}.
\newblock \bibinfo{journal}{IEEE Transactions on Software Engineering} \bibinfo{volume}{36}, \bibinfo{pages}{20--36}.
\bibitem[{Monat et~al.(2021)Monat, Ouadjaout and Min{\'e}}]{monat2021multilanguage}
\bibinfo{author}{Monat, R.}, \bibinfo{author}{Ouadjaout, A.}, \bibinfo{author}{Min{\'e}, A.}, \bibinfo{year}{2021}.
\newblock \bibinfo{title}{A multilanguage static analysis of python programs with native c extensions}, in: \bibinfo{booktitle}{Proceedings of the 28th International Static Analysis Symposium (SAS)}, \bibinfo{organization}{Springer}. pp. \bibinfo{pages}{323--345}.
\bibitem[{Overflow(2017)}]{stackoverflow}
\bibinfo{author}{Overflow, S.}, \bibinfo{year}{2017}.
\newblock \bibinfo{title}{What is the path of the loaded dll?}
\newblock \bibinfo{note}{\href{https://stackoverflow.com/questions/44453489/what-is-the-path-of-the-loaded-dll}{https://stackoverflow.com/questions/44453489/what-is-the-path-of-the-loaded-dll}}.
\bibitem[{Paszke et~al.(2019)Paszke, Gross, Massa, Lerer, Bradbury, Chanan, Killeen, Lin, Gimelshein, Antiga et~al.}]{paszke2019pytorch}
\bibinfo{author}{Paszke, A.}, \bibinfo{author}{Gross, S.}, \bibinfo{author}{Massa, F.}, \bibinfo{author}{Lerer, A.}, \bibinfo{author}{Bradbury, J.}, \bibinfo{author}{Chanan, G.}, \bibinfo{author}{Killeen, T.}, \bibinfo{author}{Lin, Z.}, \bibinfo{author}{Gimelshein, N.}, \bibinfo{author}{Antiga, L.}, et~al., \bibinfo{year}{2019}.
\newblock \bibinfo{title}{Pytorch: An imperative style, high-performance deep learning library}.
\newblock \bibinfo{journal}{Advances in Neural Information Processing Systems} \bibinfo{volume}{32}.
\bibitem[{Python(2019)}]{ast}
\bibinfo{author}{Python}, \bibinfo{year}{2019}.
\newblock \bibinfo{title}{Abstract syntax tree}.
\newblock \bibinfo{note}{\href{https://docs.python.org/3/library/ast.html\#}{https://docs.python.org/3/library/ast.html\#}}.
\bibitem[{Python(2023a)}]{python/c_api_lsd}
\bibinfo{author}{Python}, \bibinfo{year}{2023}a.
\newblock \bibinfo{title}{Extending python with c or c++}.
\newblock \bibinfo{note}{\href{https://docs.python.org/zh-cn/3/extending/extending.html}{https://docs.python.org/zh-cn/3/extending/extending.html}}.
\bibitem[{Python(2023b)}]{python/c_api_lrec}
\bibinfo{author}{Python}, \bibinfo{year}{2023}b.
\newblock \bibinfo{title}{Python/c api}.
\newblock \bibinfo{note}{\href{https://docs.python.org/3/c-api/module.html\#c.PyModule\_AddObject}{https://docs.python.org/3/c-api/module.html\#c.PyModule\_AddObject}}.
\bibitem[{Rossum(2018)}]{rossum2018extending}
\bibinfo{author}{Rossum, G.v.}, \bibinfo{year}{2018}.
\newblock \bibinfo{title}{Extending and embedding python: Release 3.6. 4}.
\bibitem[{Tokui et~al.(2015)Tokui, Oono, Hido and Clayton}]{tokui2015chainer}
\bibinfo{author}{Tokui, S.}, \bibinfo{author}{Oono, K.}, \bibinfo{author}{Hido, S.}, \bibinfo{author}{Clayton, J.}, \bibinfo{year}{2015}.
\newblock \bibinfo{title}{Chainer: a next-generation open source framework for deep learning}, in: \bibinfo{booktitle}{Proceedings of workshop on machine learning systems (LearningSys) in the twenty-ninth annual conference on neural information processing systems (NIPS)}, pp. \bibinfo{pages}{1--6}.
\bibitem[{Van~Oort et~al.(2021)Van~Oort, Cruz, Aniche and Van~Deursen}]{van2021prevalence}
\bibinfo{author}{Van~Oort, B.}, \bibinfo{author}{Cruz, L.}, \bibinfo{author}{Aniche, M.}, \bibinfo{author}{Van~Deursen, A.}, \bibinfo{year}{2021}.
\newblock \bibinfo{title}{The prevalence of code smells in machine learning projects}, in: \bibinfo{booktitle}{Proceedings of the IEEE/ACM 1st Workshop on AI Engineering-Software Engineering for AI (WAIN)}, \bibinfo{organization}{IEEE}. pp. \bibinfo{pages}{1--8}.
\bibitem[{Viera et~al.(2005)Viera, Garrett et~al.}]{viera2005understanding}
\bibinfo{author}{Viera, A.J.}, \bibinfo{author}{Garrett, J.M.}, et~al., \bibinfo{year}{2005}.
\newblock \bibinfo{title}{Understanding interobserver agreement: the kappa statistic}.
\newblock \bibinfo{journal}{Family Medicine} \bibinfo{volume}{37}, \bibinfo{pages}{360--363}.
\bibitem[{Yang et~al.(2023)Yang, Lian, Wang and Cai}]{yang2023demystifying}
\bibinfo{author}{Yang, H.}, \bibinfo{author}{Lian, W.}, \bibinfo{author}{Wang, S.}, \bibinfo{author}{Cai, H.}, \bibinfo{year}{2023}.
\newblock \bibinfo{title}{Demystifying issues, challenges, and solutions for multilingual software development}, in: \bibinfo{booktitle}{Proceedings of the IEEE/ACM 45th International Conference on Software Engineering (ICSE)}, \bibinfo{publisher}{IEEE}. pp. \bibinfo{pages}{1840--1852}.
\bibitem[{Yang et~al.(2024)Yang, Nong, Wang and Cai}]{yang2024multi}
\bibinfo{author}{Yang, H.}, \bibinfo{author}{Nong, Y.}, \bibinfo{author}{Wang, S.}, \bibinfo{author}{Cai, H.}, \bibinfo{year}{2024}.
\newblock \bibinfo{title}{Multi-language software development: Issues, challenges, and solutions}.
\newblock \bibinfo{journal}{IEEE Transactions on Software Engineering} .
\bibitem[{Yang et~al.(2022)Yang, He, Xia and Feng}]{yang2022comprehensive}
\bibinfo{author}{Yang, Y.}, \bibinfo{author}{He, T.}, \bibinfo{author}{Xia, Z.}, \bibinfo{author}{Feng, Y.}, \bibinfo{year}{2022}.
\newblock \bibinfo{title}{A comprehensive empirical study on bug characteristics of deep learning frameworks}.
\newblock \bibinfo{journal}{Information and Software Technology} \bibinfo{volume}{151}, \bibinfo{pages}{107004}.
\bibitem[{Youn et~al.(2023)Youn, Lee and Ryu}]{youn2023declarative}
\bibinfo{author}{Youn, D.}, \bibinfo{author}{Lee, S.}, \bibinfo{author}{Ryu, S.}, \bibinfo{year}{2023}.
\newblock \bibinfo{title}{Declarative static analysis for multilingual programs using codeql}.
\newblock \bibinfo{journal}{Software: Practice and Experience} .
\bibitem[{Zhang et~al.(2022)Zhang, Cruz and Van~Deursen}]{zhang2022code}
\bibinfo{author}{Zhang, H.}, \bibinfo{author}{Cruz, L.}, \bibinfo{author}{Van~Deursen, A.}, \bibinfo{year}{2022}.
\newblock \bibinfo{title}{Code smells for machine learning applications}, in: \bibinfo{booktitle}{Proceedings of the 1st International Conference on AI Engineering: Software Engineering for AI}, pp. \bibinfo{pages}{217--228}.

\end{thebibliography}

\balance










\end{sloppypar}
\end{document}